\documentclass[a4paper,11pt]{article}
\usepackage{jheppub}

\usepackage{amsmath}
\usepackage{comment}
\usepackage{graphicx}
\usepackage{tikz}
\usepackage{esvect}
\usepackage{comment}
\def\be{\begin{eqnarray}}
\def\ed{\end{eqnarray}}
\def\beq{\begin{equation}}
\def\eeq{\end{equation}}
\def\bea{\begin{eqnarray}}
\def\eea{\end{eqnarray}}
\usepackage{multicol}
\usepackage{multirow}
\usepackage[normalem]{ulem}

\begin{document}

\title{\bf \Large Probing the CP Property of ALP-photon Interactions at Future Lepton Colliders }

\author[a]{Jian-Nan Ding}
\author[b,c]{Yandong Liu}
\author[d,e]{Muyuan Song}
\affiliation[a]{Center for High Energy Physics, Peking University, Beijing 100871, China}
%\affiliation[b]{School of Physics and State Key Laboratory of Nuclear Physics and Technology, Peking University, Beijing 100871, China}
\affiliation[b]{School of Physics and Astronomy, Beijing Normal University, Beijing, 100875, China}
\affiliation[c]{Key Laboratory of Multiscale Spin Physics (Ministry of Education), Beijing Normal University, Beijing, 100875, China}
\affiliation[d]{\it\small  State Key Laboratory of Nuclear Physics and
Technology, Institute of Quantum Matter, South China Normal
University, Guangzhou 510006, China }
\affiliation[e]{\it\small Guangdong Basic Research Center of Excellence for
Structure and Fundamental Interactions of Matter, Guangdong
Provincial Key Laboratory of Nuclear Science, Guangzhou
510006, China}
\emailAdd{dingjn23@pku.edu.cn}
\emailAdd{ydliu@bnu.edu.cn} 
\emailAdd{muyuansong@m.scnu.edu.cn}

\date{\today}% It is always \today, today,

\abstract{
We investigate a charge-parity (CP) odd axion-like particle (ALP) featuring simultaneous CP-conserving ($a F_{\mu\nu}\tilde{F}^{\mu\nu}$) and CP-violating ($a F_{\mu\nu}F^{\mu\nu}$) ALP--photon interactions at future lepton colliders. 
The ALP signal is studied in the process $e^+e^- \to e^+e^- a \to e^+e^- \gamma\gamma$, where the CP structure of the interaction can be probed using the azimuthal angle difference between the final-state electrons, $\Delta\phi_{ee}$. 
We show that the projected sensitivity to the ALP–photon couplings can reach $\mathcal{O}(10^{-3})~\mathrm{TeV}^{-1}$, exceeding current constraints from the electron electric dipole moment ($e$EDM).
Because purely CP-conserving, purely CP-violating, and mixed interactions generate distinct $\Delta\phi_{ee}$ distributions, a binned likelihood analysis of this observable enables an efficient discrimination of the ALP interaction structure. 
In particular, if the CP-conserving and CP-violating couplings are comparable—as motivated by possible symmetry considerations—the interference pattern in the $\Delta\phi_{ee}$ distribution allows future lepton colliders to identify CP violation in the ALP sector once a signal is observed. 
For scenarios where the two couplings differ significantly, increasing the integrated luminosity substantially improves the sensitivity to CP-violating effects.
}

\maketitle

\section{Introduction}\label{sec:intro}

Axion-like particles (ALPs) are pseudo-Nambu-Goldstone bosons arising from the spontaneous breaking of a global symmetry analogous to Peccei-Quinn (PQ),  extending beyond the Standard Model (SM) framework~\cite{Peccei:1977hh,Peccei:1977ur,Weinberg:1977ma,Wilczek:1977pj}. Unlike the QCD axion, ALPs may originate from a broader class of ultraviolet theories--including string theory~\cite{Giannotti:2022euq,Arvanitaki:2009fg,Svrcek:2006yi}--and exhibit a wilder parameter space due to the unfixed relationship between their mass and decay constant. This flexibility also opens up new avenues for exploring phenomena such as leptogenesis or baryogenesis~\cite{Cataldi:2024bcs,Jeong:2018jqe}. Consequently, the study of ALPs has attracted substantial interdisciplinary interest across theoretical and experimental frontiers. Lighter ALPs are probed through astrophysical and cosmological signatures, beam-dump experiments, and collider-based searches targeting masses at or above the electroweak scale~\cite{AxionLimits, Dobrich:2015jyk, Bjorken:1988as, NA64:2020qwq, Dusaev:2020gxi, NA64:2021aiq, Kleban:2005rj, Mimasu:2014nea, Brivio:2017ije, Batell:2009yf, BaBar:2014zli, Belle-II:2020jti, Bao:2022onq,Polesello:2025gwj}. Notably, certain studies have incorporated unresolved anomalies, such as the $(g-2)_\mu$ discrepancy, into their ALP parameter space analyses~\cite{Brdar:2021pla, Ge:2021cjz, Liu:2022tqn, Galda:2023qjx}.

While many previous studies have focused on CP-conserving ALP interactions, recent 
work has increasingly emphasized the possibility of CP-violating ALPs~\cite{Choi:2016luu,
DiLuzio:2023lmd,Ramadan:2024vfc}. CP violation plays a central role in satisfying the 
Sakharov conditions required to explain the observed matter--antimatter asymmetry~
\cite{Sakharov:1967dj,Dine:2003ax}, making new CP-violating sectors particularly 
compelling. At low energies, precision measurements of the electron electric dipole 
moment ($e$EDM) provide some of the most sensitive probes of CP violation, with current 
limits reaching the level of $\mathcal{O}(10^{-29}\!-\!10^{-30})\,e\,\mathrm{cm}$~
\cite{ACME:2018yjb,Caldwell:2022xwj,Roussy:2022cmp}, far above the tiny Standard 
Model prediction of $<10^{-38}\,e\,\mathrm{cm}$~\cite{Pospelov:1991zt,Khriplovich:1997ga,
Pospelov:2013sca}. Experimental measurements constrain potential additional sources of CP violation beyond the SM, with ALP interactions serving as a well-motivated focus of theoretical investigation~\cite{Marciano:2016yhf,DiLuzio:2020oah,Evans:2025lzw}. Notably, $e$EDM limits restrict 
only the product of the CP-conserving and CP-violating ALP--photon couplings, leaving 
the underlying CP structure of the ALP sector undetermined. Additionally, contributions to the $e$EDM arise from indirect loop diagrams, which complicates the interpretation of the CP-violating effects. On the other hand, there is strong motivation for employing complementary high-energy probes. These probes are designed to access direct kinematic observables, enabling the disentanglement of the CP nature of ALP interactions in a straightforward manner. 

Although several studies have assessed the prospects for discovering ALPs at future
lepton colliders, including the Circular Electron 
Positron Collider (CEPC) \cite{Wang:2022ock, RebelloTeles:2023uig}, the International Linear Collider (ILC), and the Future Circular Collider (FCC-ee)~\cite{Steinberg:2021wbs, RebelloTeles:2023uig}, the collider phenomenology of simultaneous CP-conserving and CP-violating ALP–photon interactions remains largely unexplored.
In particular, the corresponding investigation raises two key questions:
(1) whether high-energy colliders can detect either ALP signals in channels where CP-conserving and CP-violating interactions contribute separately or simultaneously,
and (2) whether collider observables provide sufficient resolving power to discriminate the underlying CP structure of the ALP sector once a signal is observed.
To address these questions, we investigate how CP-violating effects modify the kinematic features of ALP production, focusing on the azimuthal angle separation of the final-state electrons in $e^+e^- \to e^+e^- a \to e^+e^- \gamma\gamma$. 
We show that differential 
distributions in $\Delta\phi_{ee}$ offer a direct probe of the ALP’s CP nature and enable 
a clear distinction between CP-conserving, CP-violating, and mixed scenarios. Our results demonstrate that future lepton colliders can achieve sensitivities of $\mathcal{O}(10^{-2})~\mathrm{TeV}^{-1}$ to individual interactions (either from CP-conserving or CP-violating couplings) and $\mathcal{O}(10^{-3})~\mathrm{TeV}^{-1}$ to coexistence of two interactions—surpassing current 
$e$EDM bounds. When both interactions are present, the interference pattern encoded in 
$\Delta\phi_{ee}$ allows the CP violation to be unambiguously identified. In addition, 
increased integrated luminosity substantially enhances the ability to resolve small 
CP-violating effects.

This paper is structured as follows. Section~\ref{sec:theory} provides an in-depth exploration of the effective Lagrangian of CP-violating ALP, encompassing both the CP-even and the CP-odd interaction terms. It juxtaposes the methodologies utilized in collider experiments, with a particular emphasis on the differential angular distribution of final-state particles, against the limitations set forth by $e$EDM studies. Following this, Section~\ref{sec:expe} outlines the phenomenological aspects associated with ALP production mechanisms. The phenomenological analyses are crafted to investigate the signatures of ALPs and ascertain the CP structures at prospective lepton colliders. The outcomes and insights garnered from these analyses are thoroughly expounded in Section~\ref{sec:result}, and we conclude in Section~\ref{sec:conclu}.

\section{Theoretical Setup of CP-violating ALPs}\label{sec:theory}

We consider an ALP that interacts with the electromagnetic
field through two operators and the relevant interactions can be
described by the following effective Lagrangian~\cite{DiLuzio:2023lmd}:
\bea
\label{equ:Lagrangian}
\mathcal{L}_{\rm ALP} \supset
\frac{1}{2}(\partial_\mu a)^2
-\frac{1}{2} m_a^2 a^2
+\frac{\tilde C_{\gamma}}{\Lambda}\, a\,F_{\mu\nu}\tilde F^{\mu\nu}
+\frac{C_{\gamma}}{\Lambda}\, a\,F_{\mu\nu}F^{\mu\nu},
\eea
where $a$ denotes the ALP field, $F_{\mu\nu}$ is the electromagnetic field-strength
tensor, and $\tilde F^{\mu\nu} = \frac{1}{2}\epsilon^{\mu\nu\sigma\rho}F_{\sigma\rho}$
is its dual. The parameter $m_a$ represents the ALP mass, while $\tilde C_\gamma$
and $C_\gamma$ characterize the strengths of the two ALP--photon interactions.
The scale $\Lambda$ denotes the cutoff associated with new physics, and we fix
$\Lambda = 1~\text{TeV}$ throughout this work for definiteness.

To identify the origin of CP violation in this effective theory and its observable consequences, it is essential to clarify the CP properties of the interaction terms in Eq.~(\ref{equ:Lagrangian}). 
In general, the CP nature of ALP--photon couplings depends not only on the electromagnetic operators themselves, but also on the CP transformation of the ALP field $a$. 
While $F_{\mu\nu}F^{\mu\nu}$ is CP-even and $F_{\mu\nu}\tilde F^{\mu\nu}$ is CP-odd, the overall CP properties of the composite operators $aF_{\mu\nu}F^{\mu\nu}$ and $aF_{\mu\nu}\tilde F^{\mu\nu}$ are determined by whether the ALP is CP-even or CP-odd. 
Take the ALP to be a pseudoscalar (CP-odd), then the operator $aF_{\mu\nu}\tilde F^{\mu\nu}$ is CP-even, while $aF_{\mu\nu}F^{\mu\nu}$ is CP-odd.
If instead ALP were a CP-even scalar, the situation would be reversed: $aF_{\mu\nu}\tilde F^{\mu\nu}$ would be CP-odd, while $aF_{\mu\nu}F^{\mu\nu}$ is CP-even. 
Consequently, neither operator alone induces observable CP violation in general. 
A genuine source of CP violation arises only when both interactions are simultaneously present, such that their interference induces CP violation at the amplitude level, which in turn manifests as CP-odd kinematic correlations in physical observables. 
This interference-based mechanism provides a direct avenue to probe the CP structure of ALP--photon interactions at high-energy colliders. 
In this work, to avoid definitional ambiguity, we adopt the convention--within the framework where the ALP is treated as a pseudoscalar--that $aF_{\mu\nu}\tilde{F}^{\mu\nu}$ corresponds to a CP-even term, while $aF_{\mu\nu}F^{\mu\nu}$ designates a CP-odd term.

\begin{figure}
    \centering    

\tikzset{every picture/.style={line width=0.75pt}} %set default line width to 0.75pt        

\begin{tikzpicture}[x=0.75pt,y=0.75pt,yscale=-1,xscale=1]
%uncomment if require: \path (0,406); %set diagram left start at 0, and has height of 406

%Straight Lines [id:da8340571609902583] 
\draw [line width=0.75]    (94.5,127) -- (156.5,159) ;
\draw [shift={(129.94,145.29)}, rotate = 207.3] [fill={rgb, 255:red, 0; green, 0; blue, 0 }  ][line width=0.08]  [draw opacity=0] (8.93,-4.29) -- (0,0) -- (8.93,4.29) -- cycle    ;
%Straight Lines [id:da9080056241043764] 
\draw [line width=0.75]    (377.5,137) -- (422.5,204) ;
\draw [shift={(402.79,174.65)}, rotate = 236.11] [fill={rgb, 255:red, 0; green, 0; blue, 0 }  ][line width=0.08]  [draw opacity=0] (8.93,-4.29) -- (0,0) -- (8.93,4.29) -- cycle    ;
%Straight Lines [id:da1676873358965527] 
\draw [line width=0.75]    (156.5,233) -- (96.5,267) ;
\draw [shift={(122.15,252.47)}, rotate = 330.46] [fill={rgb, 255:red, 0; green, 0; blue, 0 }  ][line width=0.08]  [draw opacity=0] (8.93,-4.29) -- (0,0) -- (8.93,4.29) -- cycle    ;
%Straight Lines [id:da19620600907810148] 
\draw    (156.5,159) .. controls (158.8,159.51) and (159.7,160.91) .. (159.19,163.21) .. controls (158.68,165.51) and (159.58,166.92) .. (161.88,167.43) .. controls (164.18,167.94) and (165.08,169.34) .. (164.58,171.64) .. controls (164.07,173.94) and (164.97,175.34) .. (167.27,175.85) .. controls (169.57,176.36) and (170.47,177.77) .. (169.96,180.07) .. controls (169.45,182.37) and (170.35,183.77) .. (172.65,184.28) .. controls (174.95,184.79) and (175.85,186.19) .. (175.34,188.49) .. controls (174.83,190.8) and (175.73,192.2) .. (178.04,192.71) -- (179.5,195) -- (179.5,195) ;
%Straight Lines [id:da1220144714564173] 
\draw    (179.5,195) .. controls (180.06,197.29) and (179.2,198.71) .. (176.91,199.28) .. controls (174.62,199.85) and (173.76,201.27) .. (174.32,203.56) .. controls (174.88,205.85) and (174.02,207.27) .. (171.73,207.83) .. controls (169.44,208.4) and (168.58,209.82) .. (169.14,212.11) .. controls (169.7,214.4) and (168.84,215.82) .. (166.55,216.39) .. controls (164.26,216.96) and (163.4,218.38) .. (163.97,220.67) .. controls (164.53,222.96) and (163.67,224.38) .. (161.38,224.94) .. controls (159.09,225.51) and (158.23,226.93) .. (158.79,229.22) -- (156.5,233) -- (156.5,233) ;
%Straight Lines [id:da9889242981030424] 
\draw  [dash pattern={on 4.5pt off 4.5pt}]  (179.5,195) -- (242.5,195) ;
%Straight Lines [id:da0671089309579933] 
\draw [line width=0.75]    (156.5,159) -- (242.5,126) ;
\draw [shift={(204.17,140.71)}, rotate = 159.01] [fill={rgb, 255:red, 0; green, 0; blue, 0 }  ][line width=0.08]  [draw opacity=0] (8.93,-4.29) -- (0,0) -- (8.93,4.29) -- cycle    ;
%Straight Lines [id:da22598214882739676] 
\draw [line width=0.75]    (242.5,269) -- (156.5,233) ;
\draw [shift={(194.89,249.07)}, rotate = 22.71] [fill={rgb, 255:red, 0; green, 0; blue, 0 }  ][line width=0.08]  [draw opacity=0] (8.93,-4.29) -- (0,0) -- (8.93,4.29) -- cycle    ;
%Shape: Circle [id:dp38544534205857084] 
\draw  [fill={rgb, 255:red, 0; green, 0; blue, 0 }  ,fill opacity=1 ] (174.75,194.63) .. controls (174.75,192.76) and (176.26,191.25) .. (178.13,191.25) .. controls (179.99,191.25) and (181.5,192.76) .. (181.5,194.63) .. controls (181.5,196.49) and (179.99,198) .. (178.13,198) .. controls (176.26,198) and (174.75,196.49) .. (174.75,194.63) -- cycle ;
%Straight Lines [id:da030821487775417822] 
\draw [line width=0.75]    (422.5,204) -- (382.5,271) ;
\draw [shift={(399.94,241.79)}, rotate = 300.84] [fill={rgb, 255:red, 0; green, 0; blue, 0 }  ][line width=0.08]  [draw opacity=0] (8.93,-4.29) -- (0,0) -- (8.93,4.29) -- cycle    ;
%Straight Lines [id:da4207202153536094] 
\draw    (470.5,204) .. controls (468.83,205.67) and (467.17,205.67) .. (465.5,204) .. controls (463.83,202.33) and (462.17,202.33) .. (460.5,204) .. controls (458.83,205.67) and (457.17,205.67) .. (455.5,204) .. controls (453.83,202.33) and (452.17,202.33) .. (450.5,204) .. controls (448.83,205.67) and (447.17,205.67) .. (445.5,204) .. controls (443.83,202.33) and (442.17,202.33) .. (440.5,204) .. controls (438.83,205.67) and (437.17,205.67) .. (435.5,204) .. controls (433.83,202.33) and (432.17,202.33) .. (430.5,204) .. controls (428.83,205.67) and (427.17,205.67) .. (425.5,204) -- (421.5,204) -- (421.5,204) ;
%Straight Lines [id:da6821892451519609] 
\draw    (471.5,204) .. controls (473.81,203.53) and (475.2,204.46) .. (475.66,206.77) .. controls (476.12,209.08) and (477.51,210.01) .. (479.82,209.55) .. controls (482.13,209.08) and (483.52,210.01) .. (483.98,212.32) .. controls (484.44,214.63) and (485.83,215.56) .. (488.14,215.09) .. controls (490.45,214.63) and (491.84,215.56) .. (492.3,217.87) .. controls (492.76,220.18) and (494.15,221.11) .. (496.46,220.64) .. controls (498.77,220.17) and (500.16,221.1) .. (500.62,223.41) .. controls (501.08,225.72) and (502.47,226.65) .. (504.78,226.19) .. controls (507.09,225.72) and (508.48,226.65) .. (508.94,228.96) -- (510.5,230) -- (510.5,230) ;
%Straight Lines [id:da07878220260684565] 
\draw  [dash pattern={on 4.5pt off 4.5pt}]  (472.5,204) -- (568.5,138) ;
%Straight Lines [id:da013508178202205734] 
\draw [line width=0.75]    (571.5,272) -- (510.5,230) ;
\draw [shift={(536.88,248.16)}, rotate = 34.55] [fill={rgb, 255:red, 0; green, 0; blue, 0 }  ][line width=0.08]  [draw opacity=0] (8.93,-4.29) -- (0,0) -- (8.93,4.29) -- cycle    ;
%Straight Lines [id:da63425058312954] 
\draw [line width=0.75]    (510.5,230) -- (571.5,196) ;
\draw [shift={(545.37,210.57)}, rotate = 150.87] [fill={rgb, 255:red, 0; green, 0; blue, 0 }  ][line width=0.08]  [draw opacity=0] (8.93,-4.29) -- (0,0) -- (8.93,4.29) -- cycle    ;
%Shape: Circle [id:dp12575312451189347] 
\draw  [fill={rgb, 255:red, 0; green, 0; blue, 0 }  ,fill opacity=1 ] (477.5,204) .. controls (477.55,201.8) and (475.81,200.01) .. (473.61,200.01) .. controls (471.4,200.01) and (469.58,201.8) .. (469.52,204) .. controls (469.47,206.2) and (471.21,207.99) .. (473.42,207.99) .. controls (475.62,207.99) and (477.45,206.2) .. (477.5,204) -- cycle ;

% Text Node
\draw (76,107.9) node [anchor=north west][inner sep=0.75pt]    {$e^{-}$};
% Text Node
\draw (77,267.9) node [anchor=north west][inner sep=0.75pt]    {$e^{+}$};
% Text Node
\draw (245,269.9) node [anchor=north west][inner sep=0.75pt]    {$e^{+}$};
% Text Node
\draw (246.5,105.9) node [anchor=north west][inner sep=0.75pt]    {$e^{-}$};
% Text Node
\draw (359,118.9) node [anchor=north west][inner sep=0.75pt]    {$e^{-}$};
% Text Node
\draw (575,184.9) node [anchor=north west][inner sep=0.75pt]    {$e^{-}$};
% Text Node
\draw (363,274.9) node [anchor=north west][inner sep=0.75pt]    {$e^{+}$};
% Text Node
\draw (573.5,274.4) node [anchor=north west][inner sep=0.75pt]    {$e^{+}$};
% Text Node
\draw (248,189.4) node [anchor=north west][inner sep=0.75pt]    {$a$};
% Text Node
\draw (574,121.4) node [anchor=north west][inner sep=0.75pt]    {$a$};
% Text Node
\draw (176,211.4) node [anchor=north west][inner sep=0.75pt]    {$\gamma $};
% Text Node
\draw (178,160.4) node [anchor=north west][inner sep=0.75pt]    {$\gamma $};
% Text Node
\draw (479,222.4) node [anchor=north west][inner sep=0.75pt]    {$\gamma $};
% Text Node
\draw (440,180.4) node [anchor=north west][inner sep=0.75pt]    {$\gamma $};

\end{tikzpicture}

\caption{The Feynman diagrams for $e^+e^-\to e^+e^-a$ channel. The left panel represents the ALP production via vector boson fusion, and the right panel illustrates the ALP production in $s$--channel.
  }
    \label{fig:eetoeeadiagram}
\end{figure}
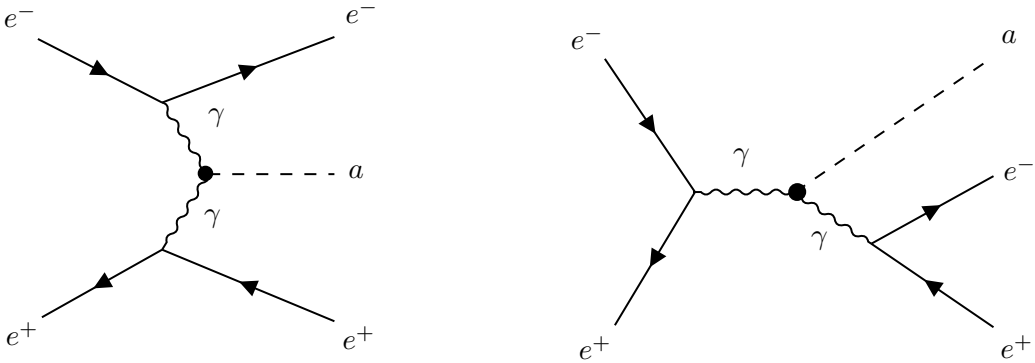

\begin{figure}
\centering
\includegraphics[width=0.50\linewidth]{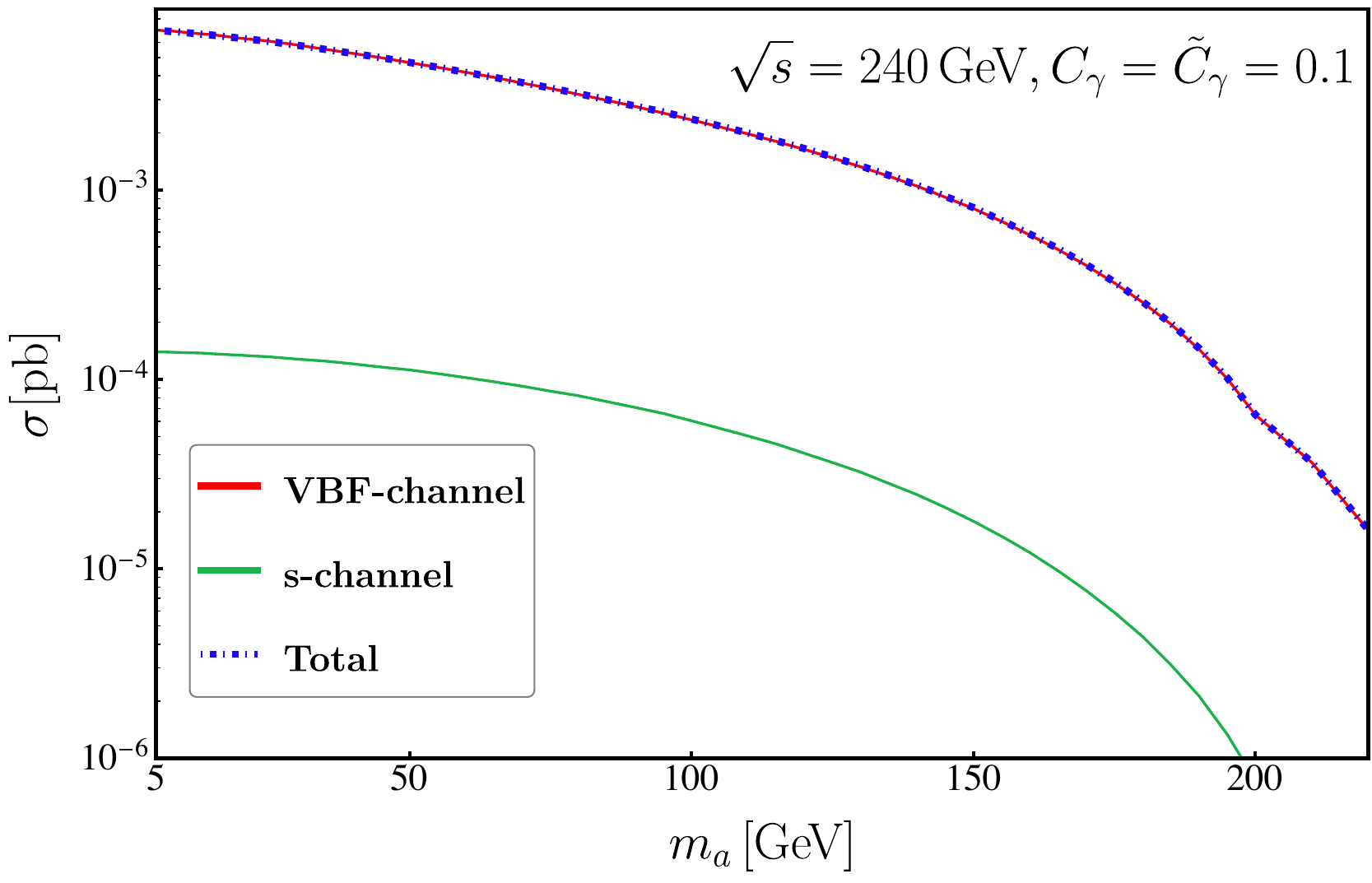}
\includegraphics[width=0.48\linewidth]{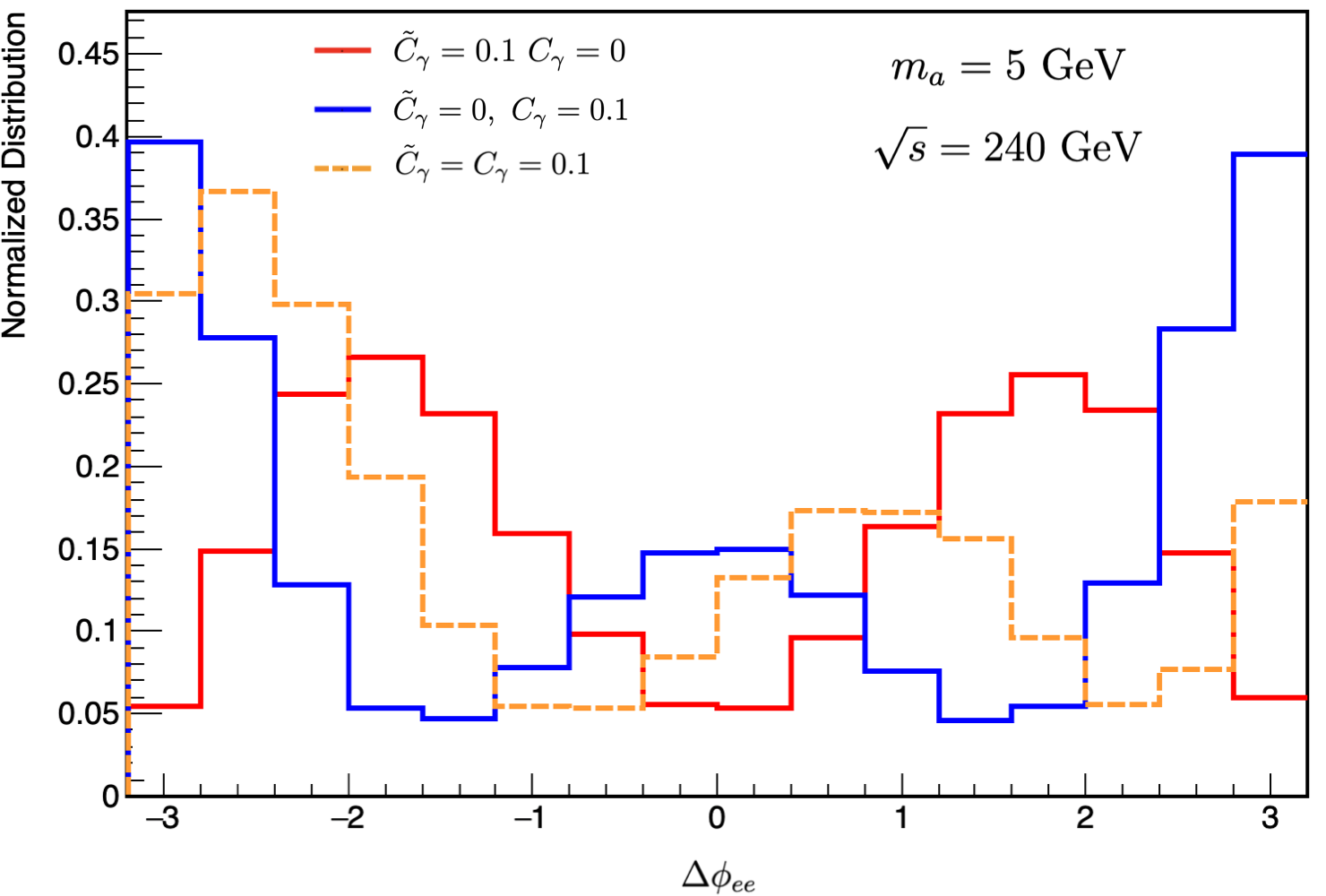}
\caption{Left panel: The cross-section for the process $\sigma (e^+ e^- \to e^+ e^- a)$ is depicted as a function of the ALP mass $m_a$ considering three distinct contributions: Vector Boson Fusion (VBF) represented in red, s-channel denoted in green, and the combined effect of all contributing diagrams shown as dot-dashed blue. The values for both $C_\gamma $ and $ \tilde{C}_\gamma$ are set to be 0.1. Right panel: Azimuthal angular difference ($\Delta\phi_{ee}$) distributions for $e^+e^-$ pairs, assuming an ALP mass $m_a = 5$~GeV.}
\label{fig:DeltaPhiee_eeAx}
\end{figure}

At low energies, CP-violating ALP interactions are tightly constrained by
precision measurements of the electron electric dipole moment ($e$EDM).
The ACME collaboration has reported a $90\%$ confidence level (C.L.) upper bound
$|d_e| < 1.1\times 10^{-29}\, e\,\text{cm}$~\cite{ACME:2018yjb}, which was
subsequently improved to $|d_e| < 4.1\times 10^{-30}\, e\,\text{cm}$ by the
JILA experiment~\cite{Roussy:2022cmp}. These measurements constrain the product
of the CP-even and CP-odd ALP--photon couplings, $|\tilde C_\gamma C_\gamma|$,
while leaving their individual values and relative CP structure undetermined.
This limitation motivates complementary probes at high-energy colliders,
where CP violation can be accessed directly through kinematic observables.

Future high-luminosity lepton colliders provide a particularly clean environment
to explore such effects. In this study, we focus on the process
$e^+e^- \to e^+e^- a$, whose representative Feynman diagrams are shown in
Fig.~\ref{fig:eetoeeadiagram}. Owing to the suppressed QCD background and well-controlled
initial state, lepton colliders allow precise measurements of ALP couplings. Using the effective Lagrangian in Eq.~(\ref{equ:Lagrangian}),
we evaluate the contributions from both vector boson fusion (VBF) and $s$-channel
production mechanisms. As illustrated in the left panel of Fig.~\ref{fig:DeltaPhiee_eeAx},
the VBF process dominates over the $s$-channel contribution across a wide range
of ALP masses, motivating our emphasis on the VBF topology in the following analysis.

The distinct CP properties of the operators $aF_{\mu\nu}\tilde F^{\mu\nu}$ and
$aF_{\mu\nu}F^{\mu\nu}$ manifest themselves in characteristic angular correlations
of the final-state leptons. Building on earlier collider studies of CP-sensitive
observables~\cite{Donoghue:1987wu,Valencia:1988it,Atwood:1994kn,Ellis:2008hq,Gronau:2011cf,Kittel:2012jrm,Hagiwara:2016zqz,Basu:2024fet,Gurkanli:2024qaf,Bigaran:2024tmp,LHCb:2025ray,Girmohanta:2023tdr}, we propose the
azimuthal angle separation between the outgoing electron and positron,
\bea
\Delta\phi_{ee} = \phi_{e^+} - \phi_{e^-},
\label{eq:delta_phi}
\eea
as a sensitive probe of the ALP CP structure. 
Under a CP transformation, $\Delta\phi_{ee} \to -\Delta\phi_{ee}$; therefore, any antisymmetric component of the distribution $d\sigma/d\Delta\phi_{ee}$ constitutes a genuine CP-odd observable.
As shown in the right panel of
Fig.~\ref{fig:DeltaPhiee_eeAx}, different combinations of CP-even and CP-odd couplings
lead to distinct $\Delta\phi_{ee}$ distributions. In particular, the interference
between the two operators induces an asymmetric component in $\Delta\phi_{ee}$,
providing a direct signature of CP violation. These features establish
$\Delta\phi_{ee}$ as a robust and experimentally accessible observable for
discriminating the CP nature of ALP--photon interactions at future lepton colliders.

\section{Probing ALPs at Future Lepton Colliders}\label{sec:expe}

\subsection{Phenomenological Analysis at Future Lepton Colliders}

To enable collider phenomenology studies, we implement the ALP interaction model in FeynRules \cite{Alloul:2013bka,Christensen:2008py}, generating Universal FeynRules Output (UFO) modules for MadGraph \cite{Alwall:2014hca}, which calculates leading-order matrix elements for both signal and background processes. 
The parton-level events are interfaced with Pythia8 \cite{Sjostrand:2007gs,Sjostrand:2014zea} to simulate parton showering, hadronization, and underlying event effects. 
For detector-level analysis, we employ Delphes \cite{deFavereau:2013fsa} with a parameterized detector response tailored to the CEPC experiment setup, modeling detector resolution and acceptance effects. 
To ensure precise reconstruction of ALPs, we account for the photon energy resolution, which is parameterized as follow \cite{Ai:2024nmn,Wang:2022nrm}:
\begin{equation}
    \frac{\Delta E_\gamma}{E_\gamma}=\frac{3\%}{\sqrt{E_\gamma\,(\rm GeV)}}\oplus 1\%.
\end{equation}

\begin{figure}
\centering
\includegraphics[width=0.463\linewidth]{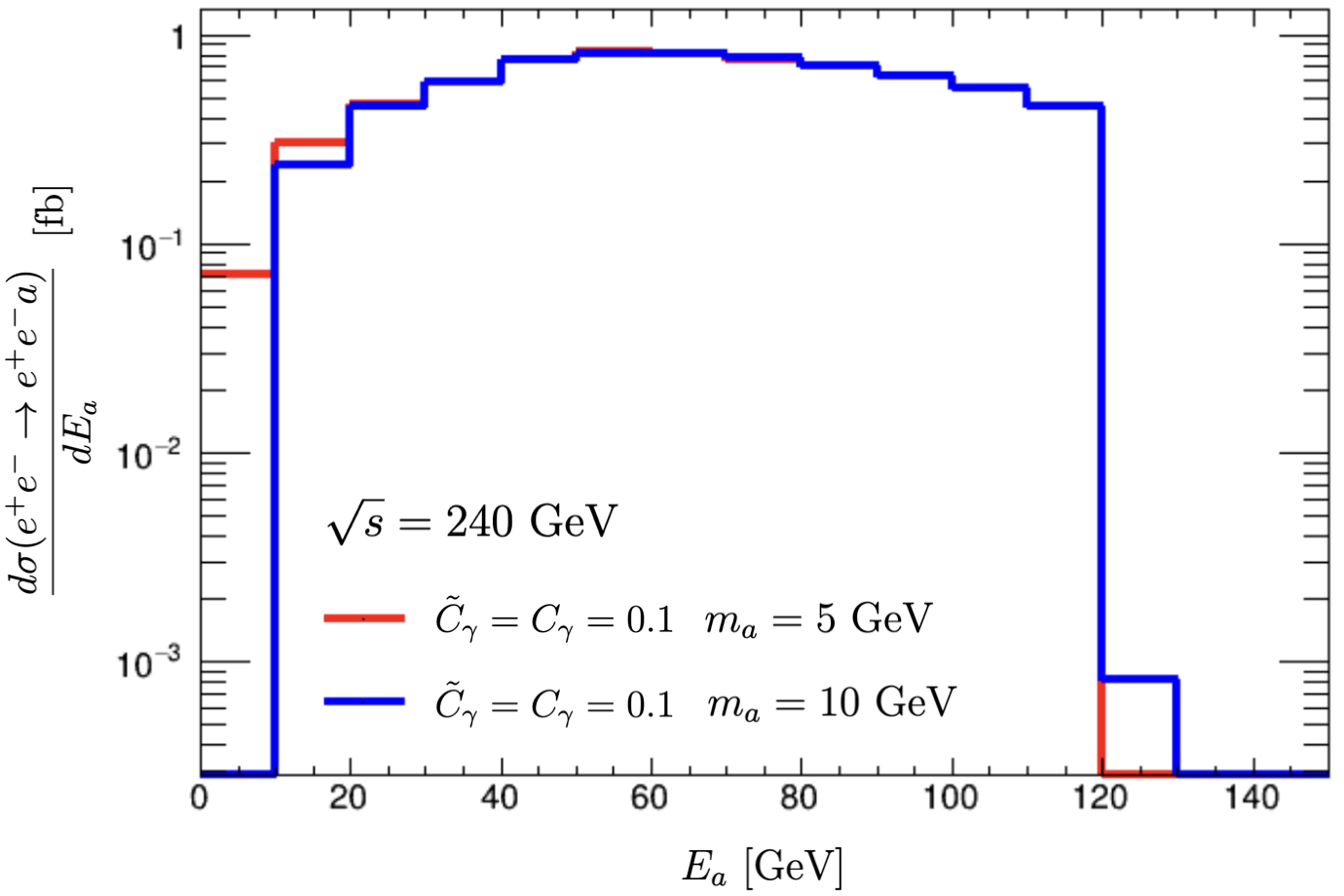}
\includegraphics[width=0.473\linewidth]{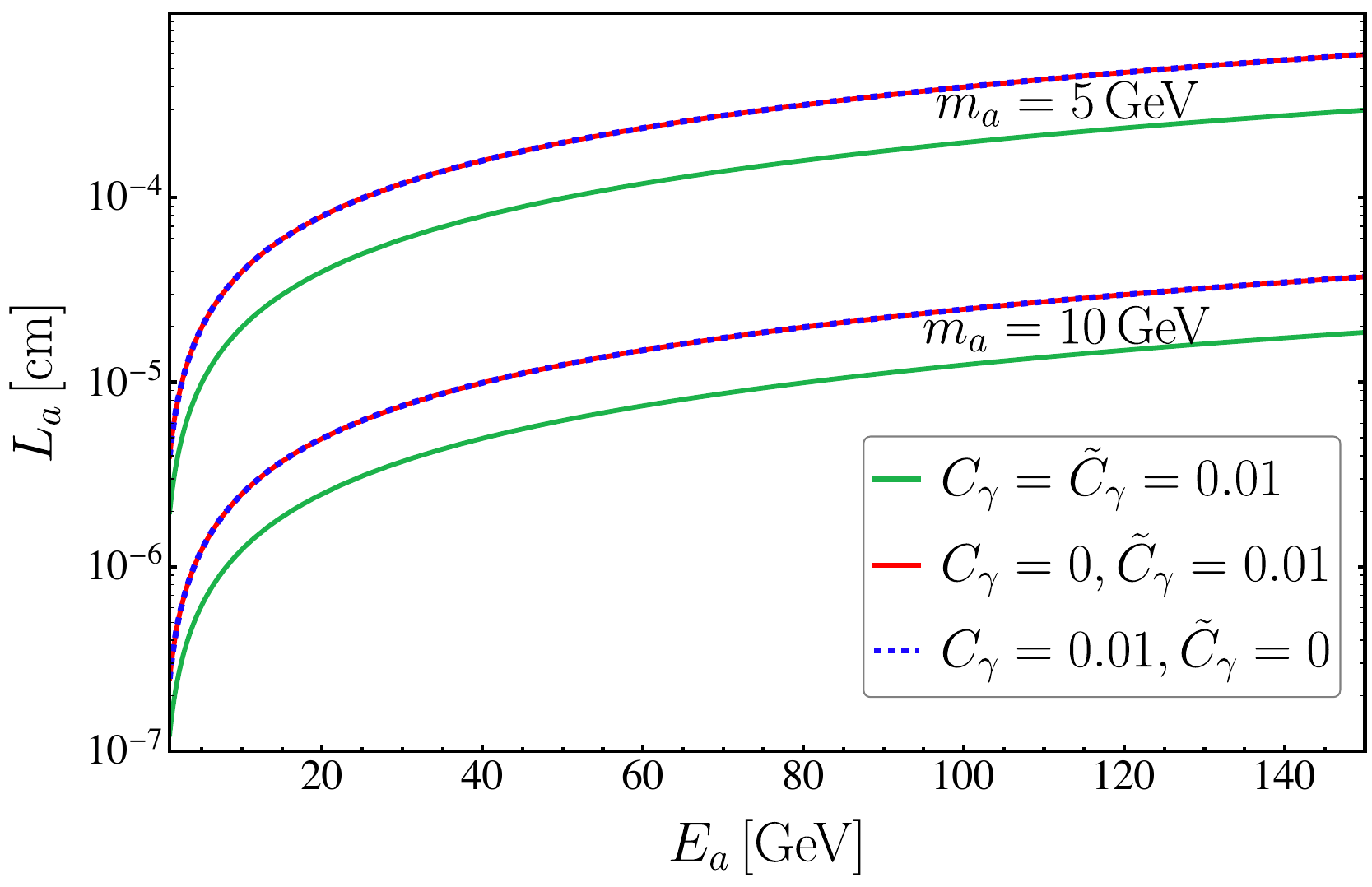}
\caption{The energy distribution and decay length for the ALPs at future lepton colliders with $\sqrt{s}=240$ GeV. 
}
\label{fig:decaylengthenergy}
\end{figure}

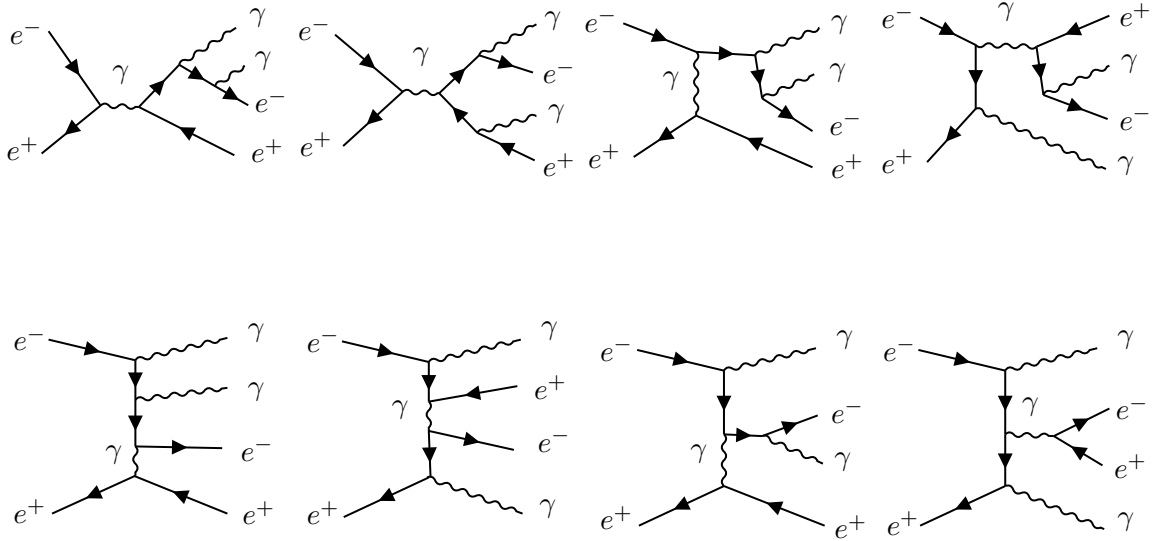
\begin{figure}
    \centering

\tikzset{every picture/.style={line width=0.75pt}} %set default line width to 0.75pt        

\begin{tikzpicture}[x=0.65pt,y=0.65pt,yscale=-1,xscale=1]
%uncomment if require: \path (0,406); %set diagram left start at 0, and has height of 406

%Straight Lines [id:da0374417952210887] 
\draw    (50,61.5) -- (79.98,104.67) ;
\draw [shift={(67.84,87.19)}, rotate = 235.22] [fill={rgb, 255:red, 0; green, 0; blue, 0 }  ][line width=0.08]  [draw opacity=0] (8.93,-4.29) -- (0,0) -- (8.93,4.29) -- cycle    ;
%Straight Lines [id:da42338696714360413] 
\draw    (79.98,104.67) -- (46,132.5) ;
\draw [shift={(59.12,121.75)}, rotate = 320.68] [fill={rgb, 255:red, 0; green, 0; blue, 0 }  ][line width=0.08]  [draw opacity=0] (8.93,-4.29) -- (0,0) -- (8.93,4.29) -- cycle    ;
%Straight Lines [id:da24716235271685405] 
\draw    (79.98,104.67) .. controls (81.71,103.07) and (83.38,103.13) .. (84.98,104.86) .. controls (86.59,106.59) and (88.25,106.65) .. (89.98,105.04) .. controls (91.71,103.44) and (93.37,103.5) .. (94.97,105.23) .. controls (96.58,106.96) and (98.24,107.02) .. (99.97,105.42) -- (102,105.5) -- (102,105.5) ;
%Straight Lines [id:da7282007982712423] 
\draw    (102,105.5) -- (125.26,81) ;
\draw [shift={(117.07,89.62)}, rotate = 133.51] [fill={rgb, 255:red, 0; green, 0; blue, 0 }  ][line width=0.08]  [draw opacity=0] (8.93,-4.29) -- (0,0) -- (8.93,4.29) -- cycle    ;
%Straight Lines [id:da6319066924880573] 
\draw    (125.26,81) -- (146,93.5) ;
\draw [shift={(139.91,89.83)}, rotate = 211.08] [fill={rgb, 255:red, 0; green, 0; blue, 0 }  ][line width=0.08]  [draw opacity=0] (8.93,-4.29) -- (0,0) -- (8.93,4.29) -- cycle    ;
%Straight Lines [id:da22235071904761894] 
\draw    (146,93.5) -- (165,104.5) ;
\draw [shift={(159.83,101.51)}, rotate = 210.07] [fill={rgb, 255:red, 0; green, 0; blue, 0 }  ][line width=0.08]  [draw opacity=0] (8.93,-4.29) -- (0,0) -- (8.93,4.29) -- cycle    ;
%Straight Lines [id:da03813130491547356] 
\draw    (125.26,81) .. controls (125.74,78.69) and (127.13,77.78) .. (129.44,78.26) .. controls (131.75,78.73) and (133.14,77.82) .. (133.62,75.51) .. controls (134.1,73.2) and (135.49,72.29) .. (137.8,72.77) .. controls (140.11,73.24) and (141.5,72.33) .. (141.98,70.02) .. controls (142.46,67.71) and (143.85,66.8) .. (146.16,67.28) .. controls (148.47,67.75) and (149.86,66.84) .. (150.34,64.53) .. controls (150.82,62.22) and (152.21,61.31) .. (154.52,61.79) -- (158,59.5) -- (158,59.5) ;
%Straight Lines [id:da2760235918620766] 
\draw    (146,93.5) .. controls (146.4,91.18) and (147.76,90.22) .. (150.08,90.62) .. controls (152.41,91.02) and (153.77,90.06) .. (154.17,87.73) .. controls (154.57,85.41) and (155.93,84.45) .. (158.25,84.85) .. controls (160.57,85.25) and (161.93,84.29) .. (162.34,81.97) -- (163,81.5) -- (163,81.5) ;
%Straight Lines [id:da47721883748934313] 
\draw    (157,133.5) -- (102,105.5) ;
\draw [shift={(125.04,117.23)}, rotate = 26.98] [fill={rgb, 255:red, 0; green, 0; blue, 0 }  ][line width=0.08]  [draw opacity=0] (8.93,-4.29) -- (0,0) -- (8.93,4.29) -- cycle    ;
%Straight Lines [id:da6463065127057286] 
\draw    (215,63.5) -- (253.98,96.67) ;
\draw [shift={(238.3,83.32)}, rotate = 220.39] [fill={rgb, 255:red, 0; green, 0; blue, 0 }  ][line width=0.08]  [draw opacity=0] (8.93,-4.29) -- (0,0) -- (8.93,4.29) -- cycle    ;
%Straight Lines [id:da8183838933607616] 
\draw    (253.98,96.67) -- (219.65,128.17) ;
\draw [shift={(233.13,115.8)}, rotate = 317.47] [fill={rgb, 255:red, 0; green, 0; blue, 0 }  ][line width=0.08]  [draw opacity=0] (8.93,-4.29) -- (0,0) -- (8.93,4.29) -- cycle    ;
%Straight Lines [id:da23504987771193298] 
\draw    (253.98,96.67) .. controls (255.71,95.07) and (257.38,95.13) .. (258.98,96.86) .. controls (260.58,98.59) and (262.25,98.66) .. (263.98,97.06) .. controls (265.71,95.46) and (267.38,95.53) .. (268.97,97.26) .. controls (270.57,98.99) and (272.24,99.06) .. (273.97,97.46) -- (275,97.5) -- (275,97.5) ;
%Straight Lines [id:da851474309500073] 
\draw    (275,97.5) -- (298,75.5) ;
\draw [shift={(290.11,83.04)}, rotate = 136.27] [fill={rgb, 255:red, 0; green, 0; blue, 0 }  ][line width=0.08]  [draw opacity=0] (8.93,-4.29) -- (0,0) -- (8.93,4.29) -- cycle    ;
%Straight Lines [id:da49366934398026574] 
\draw    (330,136.5) -- (297,120.5) ;
\draw [shift={(309,126.32)}, rotate = 25.87] [fill={rgb, 255:red, 0; green, 0; blue, 0 }  ][line width=0.08]  [draw opacity=0] (8.93,-4.29) -- (0,0) -- (8.93,4.29) -- cycle    ;
%Straight Lines [id:da7324097291061854] 
\draw    (298,75.5) -- (329,85.5) ;
\draw [shift={(318.26,82.04)}, rotate = 197.88] [fill={rgb, 255:red, 0; green, 0; blue, 0 }  ][line width=0.08]  [draw opacity=0] (8.93,-4.29) -- (0,0) -- (8.93,4.29) -- cycle    ;
%Straight Lines [id:da4108148423063295] 
\draw    (298,75.5) .. controls (298.67,73.24) and (300.13,72.44) .. (302.39,73.11) .. controls (304.65,73.77) and (306.11,72.97) .. (306.78,70.71) .. controls (307.45,68.45) and (308.91,67.65) .. (311.17,68.32) .. controls (313.43,68.98) and (314.89,68.18) .. (315.56,65.92) .. controls (316.23,63.66) and (317.69,62.86) .. (319.95,63.53) .. controls (322.21,64.19) and (323.67,63.39) .. (324.34,61.13) .. controls (325.01,58.87) and (326.47,58.07) .. (328.73,58.74) -- (331,57.5) -- (331,57.5) ;
%Straight Lines [id:da45657189502811313] 
\draw    (297,120.5) .. controls (298.13,118.43) and (299.73,117.96) .. (301.8,119.09) .. controls (303.87,120.22) and (305.46,119.75) .. (306.59,117.68) .. controls (307.72,115.61) and (309.32,115.14) .. (311.39,116.27) .. controls (313.46,117.4) and (315.06,116.93) .. (316.19,114.86) .. controls (317.32,112.79) and (318.91,112.32) .. (320.98,113.45) .. controls (323.05,114.58) and (324.65,114.11) .. (325.78,112.04) .. controls (326.91,109.97) and (328.51,109.49) .. (330.58,110.62) -- (331,110.5) -- (331,110.5) ;
%Straight Lines [id:da38053458771983684] 
\draw    (297,120.5) -- (275,97.5) ;
\draw [shift={(282.54,105.39)}, rotate = 46.27] [fill={rgb, 255:red, 0; green, 0; blue, 0 }  ][line width=0.08]  [draw opacity=0] (8.93,-4.29) -- (0,0) -- (8.93,4.29) -- cycle    ;
%Straight Lines [id:da3027912998338491] 
\draw    (381,56.92) -- (423,73.5) ;
\draw [shift={(406.65,67.04)}, rotate = 201.55] [fill={rgb, 255:red, 0; green, 0; blue, 0 }  ][line width=0.08]  [draw opacity=0] (8.93,-4.29) -- (0,0) -- (8.93,4.29) -- cycle    ;
%Straight Lines [id:da37886935230720376] 
\draw    (423,110.5) -- (387,134.5) ;
\draw [shift={(400.84,125.27)}, rotate = 326.31] [fill={rgb, 255:red, 0; green, 0; blue, 0 }  ][line width=0.08]  [draw opacity=0] (8.93,-4.29) -- (0,0) -- (8.93,4.29) -- cycle    ;
%Straight Lines [id:da2992494562161425] 
\draw    (423,73.5) .. controls (424.67,75.17) and (424.67,76.83) .. (423,78.5) .. controls (421.33,80.17) and (421.33,81.83) .. (423,83.5) .. controls (424.67,85.17) and (424.67,86.83) .. (423,88.5) .. controls (421.33,90.17) and (421.33,91.83) .. (423,93.5) .. controls (424.67,95.17) and (424.67,96.83) .. (423,98.5) .. controls (421.33,100.17) and (421.33,101.83) .. (423,103.5) .. controls (424.67,105.17) and (424.67,106.83) .. (423,108.5) -- (423,110.5) -- (423,110.5) ;
%Straight Lines [id:da03579809018275426] 
\draw    (423,73.5) -- (457,75.5) ;
\draw [shift={(444.99,74.79)}, rotate = 183.37] [fill={rgb, 255:red, 0; green, 0; blue, 0 }  ][line width=0.08]  [draw opacity=0] (8.93,-4.29) -- (0,0) -- (8.93,4.29) -- cycle    ;
%Straight Lines [id:da9484760423638167] 
\draw    (490,140.5) -- (423,110.5) ;
\draw [shift={(451.94,123.46)}, rotate = 24.12] [fill={rgb, 255:red, 0; green, 0; blue, 0 }  ][line width=0.08]  [draw opacity=0] (8.93,-4.29) -- (0,0) -- (8.93,4.29) -- cycle    ;
%Straight Lines [id:da2575167091738678] 
\draw    (461,100.5) -- (490,119.5) ;
\draw [shift={(479.68,112.74)}, rotate = 213.23] [fill={rgb, 255:red, 0; green, 0; blue, 0 }  ][line width=0.08]  [draw opacity=0] (8.93,-4.29) -- (0,0) -- (8.93,4.29) -- cycle    ;
%Straight Lines [id:da5769651087765328] 
\draw    (457,75.5) .. controls (457.87,73.31) and (459.4,72.65) .. (461.59,73.52) .. controls (463.78,74.39) and (465.31,73.72) .. (466.18,71.53) .. controls (467.05,69.34) and (468.58,68.68) .. (470.77,69.55) .. controls (472.96,70.42) and (474.49,69.75) .. (475.36,67.56) .. controls (476.23,65.37) and (477.76,64.71) .. (479.95,65.58) .. controls (482.14,66.45) and (483.67,65.78) .. (484.54,63.59) .. controls (485.41,61.4) and (486.93,60.74) .. (489.12,61.61) .. controls (491.31,62.48) and (492.84,61.81) .. (493.71,59.62) -- (494,59.5) -- (494,59.5) ;
%Straight Lines [id:da8754415746151973] 
\draw    (461,100.5) .. controls (461.81,98.29) and (463.32,97.58) .. (465.53,98.39) .. controls (467.74,99.19) and (469.25,98.48) .. (470.06,96.27) .. controls (470.87,94.06) and (472.38,93.35) .. (474.59,94.16) .. controls (476.8,94.96) and (478.31,94.25) .. (479.12,92.04) .. controls (479.93,89.83) and (481.44,89.12) .. (483.65,89.93) .. controls (485.87,90.74) and (487.38,90.03) .. (488.19,87.81) -- (491,86.5) -- (491,86.5) ;
%Straight Lines [id:da09357898140296328] 
\draw    (457,75.5) -- (461,100.5) ;
\draw [shift={(459.79,92.94)}, rotate = 260.91] [fill={rgb, 255:red, 0; green, 0; blue, 0 }  ][line width=0.08]  [draw opacity=0] (8.93,-4.29) -- (0,0) -- (8.93,4.29) -- cycle    ;
%Straight Lines [id:da17390126625673408] 
\draw    (552,53.5) -- (584,69.5) ;
\draw [shift={(572.47,63.74)}, rotate = 206.57] [fill={rgb, 255:red, 0; green, 0; blue, 0 }  ][line width=0.08]  [draw opacity=0] (8.93,-4.29) -- (0,0) -- (8.93,4.29) -- cycle    ;
%Straight Lines [id:da9713480245034073] 
\draw    (584,106.5) -- (556,137.5) ;
\draw [shift={(566.65,125.71)}, rotate = 312.09] [fill={rgb, 255:red, 0; green, 0; blue, 0 }  ][line width=0.08]  [draw opacity=0] (8.93,-4.29) -- (0,0) -- (8.93,4.29) -- cycle    ;
%Straight Lines [id:da3359566341089508] 
\draw    (619,70.5) .. controls (617.29,72.12) and (615.62,72.07) .. (614,70.36) .. controls (612.38,68.65) and (610.71,68.6) .. (609,70.21) .. controls (607.29,71.83) and (605.62,71.78) .. (604.01,70.07) .. controls (602.39,68.36) and (600.72,68.31) .. (599.01,69.93) .. controls (597.3,71.55) and (595.63,71.5) .. (594.01,69.79) .. controls (592.39,68.08) and (590.72,68.03) .. (589.01,69.64) .. controls (587.3,71.26) and (585.63,71.21) .. (584.01,69.5) -- (584,69.5) -- (584,69.5) ;
%Straight Lines [id:da3251528728694817] 
\draw    (584,69.5) -- (584,106.5) ;
\draw [shift={(584,93)}, rotate = 270] [fill={rgb, 255:red, 0; green, 0; blue, 0 }  ][line width=0.08]  [draw opacity=0] (8.93,-4.29) -- (0,0) -- (8.93,4.29) -- cycle    ;
%Straight Lines [id:da2035302152634847] 
\draw    (661,52.5) -- (619,70.5) ;
\draw [shift={(635.4,63.47)}, rotate = 336.8] [fill={rgb, 255:red, 0; green, 0; blue, 0 }  ][line width=0.08]  [draw opacity=0] (8.93,-4.29) -- (0,0) -- (8.93,4.29) -- cycle    ;
%Straight Lines [id:da4293557455781074] 
\draw    (623,98.5) -- (660,112.5) ;
\draw [shift={(646.18,107.27)}, rotate = 200.73] [fill={rgb, 255:red, 0; green, 0; blue, 0 }  ][line width=0.08]  [draw opacity=0] (8.93,-4.29) -- (0,0) -- (8.93,4.29) -- cycle    ;
%Straight Lines [id:da6705627377834804] 
\draw    (623,98.5) .. controls (623.84,96.3) and (625.36,95.62) .. (627.56,96.46) .. controls (629.76,97.3) and (631.28,96.62) .. (632.13,94.42) .. controls (632.96,92.21) and (634.48,91.53) .. (636.69,92.37) .. controls (638.89,93.21) and (640.41,92.53) .. (641.26,90.33) .. controls (642.1,88.13) and (643.62,87.45) .. (645.82,88.29) .. controls (648.02,89.13) and (649.54,88.45) .. (650.38,86.25) .. controls (651.23,84.05) and (652.75,83.37) .. (654.95,84.21) .. controls (657.15,85.05) and (658.67,84.37) .. (659.51,82.17) -- (661,81.5) -- (661,81.5) ;
%Straight Lines [id:da6938596330057157] 
\draw    (584,106.5) .. controls (586.21,105.69) and (587.72,106.4) .. (588.53,108.61) .. controls (589.34,110.82) and (590.85,111.53) .. (593.06,110.73) .. controls (595.27,109.92) and (596.78,110.63) .. (597.59,112.84) .. controls (598.4,115.05) and (599.91,115.76) .. (602.12,114.96) .. controls (604.33,114.15) and (605.84,114.86) .. (606.65,117.07) .. controls (607.46,119.29) and (608.97,120) .. (611.19,119.19) .. controls (613.4,118.38) and (614.91,119.09) .. (615.72,121.3) .. controls (616.53,123.51) and (618.04,124.22) .. (620.25,123.42) .. controls (622.46,122.61) and (623.97,123.32) .. (624.78,125.53) .. controls (625.59,127.74) and (627.1,128.45) .. (629.31,127.64) .. controls (631.52,126.84) and (633.03,127.55) .. (633.84,129.76) .. controls (634.65,131.97) and (636.16,132.68) .. (638.37,131.87) .. controls (640.58,131.07) and (642.09,131.78) .. (642.9,133.99) .. controls (643.71,136.2) and (645.22,136.91) .. (647.43,136.1) .. controls (649.64,135.3) and (651.15,136.01) .. (651.96,138.22) .. controls (652.77,140.43) and (654.28,141.14) .. (656.49,140.33) -- (659,141.5) -- (659,141.5) ;
%Straight Lines [id:da28054635126472216] 
\draw    (619,70.5) -- (623,98.5) ;
\draw [shift={(621.71,89.45)}, rotate = 261.87] [fill={rgb, 255:red, 0; green, 0; blue, 0 }  ][line width=0.08]  [draw opacity=0] (8.93,-4.29) -- (0,0) -- (8.93,4.29) -- cycle    ;
%Straight Lines [id:da5724642039980474] 
\draw    (50,240.5) -- (100,252.5) ;
\draw [shift={(79.86,247.67)}, rotate = 193.5] [fill={rgb, 255:red, 0; green, 0; blue, 0 }  ][line width=0.08]  [draw opacity=0] (8.93,-4.29) -- (0,0) -- (8.93,4.29) -- cycle    ;
%Straight Lines [id:da7836574036829985] 
\draw    (100,319.5) -- (52,341.5) ;
\draw [shift={(71.45,332.58)}, rotate = 335.38] [fill={rgb, 255:red, 0; green, 0; blue, 0 }  ][line width=0.08]  [draw opacity=0] (8.93,-4.29) -- (0,0) -- (8.93,4.29) -- cycle    ;
%Straight Lines [id:da4591208917992834] 
\draw    (100,302.5) .. controls (101.67,304.17) and (101.67,305.83) .. (100,307.5) .. controls (98.33,309.17) and (98.33,310.83) .. (100,312.5) .. controls (101.67,314.17) and (101.67,315.83) .. (100,317.5) -- (100,319.5) -- (100,319.5) ;
%Straight Lines [id:da8364012739822156] 
\draw    (100,252.5) -- (100,274.5) ;
\draw [shift={(100,268.5)}, rotate = 270] [fill={rgb, 255:red, 0; green, 0; blue, 0 }  ][line width=0.08]  [draw opacity=0] (8.93,-4.29) -- (0,0) -- (8.93,4.29) -- cycle    ;
%Straight Lines [id:da8660928598519997] 
\draw    (153,341.5) -- (99,319.5) ;
\draw [shift={(121.37,328.61)}, rotate = 22.17] [fill={rgb, 255:red, 0; green, 0; blue, 0 }  ][line width=0.08]  [draw opacity=0] (8.93,-4.29) -- (0,0) -- (8.93,4.29) -- cycle    ;
%Straight Lines [id:da04433536576052999] 
\draw    (100,302.5) -- (150,303.5) ;
\draw [shift={(130,303.1)}, rotate = 181.15] [fill={rgb, 255:red, 0; green, 0; blue, 0 }  ][line width=0.08]  [draw opacity=0] (8.93,-4.29) -- (0,0) -- (8.93,4.29) -- cycle    ;
%Straight Lines [id:da7738283951624089] 
\draw    (100,252.5) .. controls (101.23,250.49) and (102.85,250.09) .. (104.86,251.31) .. controls (106.87,252.53) and (108.49,252.13) .. (109.71,250.12) .. controls (110.94,248.11) and (112.56,247.71) .. (114.57,248.93) .. controls (116.58,250.15) and (118.2,249.75) .. (119.42,247.74) .. controls (120.64,245.72) and (122.26,245.32) .. (124.28,246.54) .. controls (126.29,247.76) and (127.91,247.36) .. (129.14,245.35) .. controls (130.36,243.34) and (131.98,242.94) .. (133.99,244.16) .. controls (136,245.38) and (137.62,244.98) .. (138.85,242.97) .. controls (140.07,240.96) and (141.69,240.56) .. (143.7,241.78) .. controls (145.71,243) and (147.33,242.6) .. (148.56,240.59) -- (153,239.5) -- (153,239.5) ;
%Straight Lines [id:da44108030091373807] 
\draw    (100,275.5) .. controls (101.44,273.63) and (103.09,273.42) .. (104.96,274.85) .. controls (106.83,276.28) and (108.48,276.06) .. (109.91,274.19) .. controls (111.35,272.32) and (113,272.11) .. (114.87,273.54) .. controls (116.74,274.97) and (118.4,274.75) .. (119.83,272.88) .. controls (121.26,271.01) and (122.91,270.8) .. (124.78,272.23) .. controls (126.65,273.66) and (128.31,273.44) .. (129.74,271.57) .. controls (131.18,269.7) and (132.83,269.49) .. (134.7,270.92) .. controls (136.57,272.35) and (138.23,272.13) .. (139.66,270.26) .. controls (141.09,268.39) and (142.74,268.18) .. (144.61,269.61) .. controls (146.48,271.04) and (148.14,270.82) .. (149.57,268.95) -- (153,268.5) -- (153,268.5) ;
%Straight Lines [id:da28524506090383106] 
\draw    (100,274.5) -- (100,302.5) ;
\draw [shift={(100,293.5)}, rotate = 270] [fill={rgb, 255:red, 0; green, 0; blue, 0 }  ][line width=0.08]  [draw opacity=0] (8.93,-4.29) -- (0,0) -- (8.93,4.29) -- cycle    ;
%Straight Lines [id:da4572347505315175] 
\draw    (219,241.5) -- (269,253.5) ;
\draw [shift={(248.86,248.67)}, rotate = 193.5] [fill={rgb, 255:red, 0; green, 0; blue, 0 }  ][line width=0.08]  [draw opacity=0] (8.93,-4.29) -- (0,0) -- (8.93,4.29) -- cycle    ;
%Straight Lines [id:da6649593498500476] 
\draw    (269,320.5) -- (220,343.5) ;
\draw [shift={(239.97,334.12)}, rotate = 334.86] [fill={rgb, 255:red, 0; green, 0; blue, 0 }  ][line width=0.08]  [draw opacity=0] (8.93,-4.29) -- (0,0) -- (8.93,4.29) -- cycle    ;
%Straight Lines [id:da7418901188566308] 
\draw    (269,276.5) .. controls (270.67,278.17) and (270.67,279.83) .. (269,281.5) .. controls (267.33,283.17) and (267.33,284.83) .. (269,286.5) .. controls (270.67,288.17) and (270.67,289.83) .. (269,291.5) -- (269,293.5) -- (269,293.5) ;
%Straight Lines [id:da45845346271286314] 
\draw    (269,253.5) -- (269,276.5) ;
\draw [shift={(269,270)}, rotate = 270] [fill={rgb, 255:red, 0; green, 0; blue, 0 }  ][line width=0.08]  [draw opacity=0] (8.93,-4.29) -- (0,0) -- (8.93,4.29) -- cycle    ;
%Straight Lines [id:da29562586691452886] 
\draw    (320,267.5) -- (269,276.5) ;
\draw [shift={(289.58,272.87)}, rotate = 349.99] [fill={rgb, 255:red, 0; green, 0; blue, 0 }  ][line width=0.08]  [draw opacity=0] (8.93,-4.29) -- (0,0) -- (8.93,4.29) -- cycle    ;
%Straight Lines [id:da39158255197390146] 
\draw    (269,293.5) -- (318,304.5) ;
\draw [shift={(298.38,300.1)}, rotate = 192.65] [fill={rgb, 255:red, 0; green, 0; blue, 0 }  ][line width=0.08]  [draw opacity=0] (8.93,-4.29) -- (0,0) -- (8.93,4.29) -- cycle    ;
%Straight Lines [id:da7910777295348811] 
\draw    (269,253.5) .. controls (270.23,251.49) and (271.85,251.09) .. (273.86,252.31) .. controls (275.87,253.53) and (277.49,253.13) .. (278.71,251.12) .. controls (279.94,249.11) and (281.56,248.71) .. (283.57,249.93) .. controls (285.58,251.15) and (287.2,250.75) .. (288.42,248.74) .. controls (289.64,246.72) and (291.26,246.32) .. (293.28,247.54) .. controls (295.29,248.76) and (296.91,248.36) .. (298.14,246.35) .. controls (299.36,244.34) and (300.98,243.94) .. (302.99,245.16) .. controls (305,246.38) and (306.62,245.98) .. (307.85,243.97) .. controls (309.07,241.96) and (310.69,241.56) .. (312.7,242.78) .. controls (314.71,244) and (316.33,243.6) .. (317.56,241.59) -- (322,240.5) -- (322,240.5) ;
%Straight Lines [id:da31047056696928843] 
\draw    (268,320.5) .. controls (270.15,319.53) and (271.71,320.13) .. (272.67,322.28) .. controls (273.63,324.43) and (275.19,325.03) .. (277.34,324.07) .. controls (279.49,323.1) and (281.05,323.7) .. (282.01,325.85) .. controls (282.97,328) and (284.53,328.6) .. (286.68,327.63) .. controls (288.83,326.67) and (290.39,327.27) .. (291.36,329.42) .. controls (292.32,331.57) and (293.88,332.17) .. (296.03,331.2) .. controls (298.18,330.23) and (299.74,330.83) .. (300.7,332.98) .. controls (301.66,335.13) and (303.22,335.73) .. (305.37,334.77) .. controls (307.52,333.8) and (309.08,334.4) .. (310.04,336.55) .. controls (311,338.7) and (312.56,339.3) .. (314.71,338.34) .. controls (316.86,337.37) and (318.42,337.97) .. (319.38,340.12) -- (323,341.5) -- (323,341.5) ;
%Straight Lines [id:da5753232346447628] 
\draw    (269,293.5) -- (270,320.5) ;
\draw [shift={(269.69,312)}, rotate = 267.88] [fill={rgb, 255:red, 0; green, 0; blue, 0 }  ][line width=0.08]  [draw opacity=0] (8.93,-4.29) -- (0,0) -- (8.93,4.29) -- cycle    ;
%Straight Lines [id:da558669351688902] 
\draw    (389,246.5) -- (439,258.5) ;
\draw [shift={(418.86,253.67)}, rotate = 193.5] [fill={rgb, 255:red, 0; green, 0; blue, 0 }  ][line width=0.08]  [draw opacity=0] (8.93,-4.29) -- (0,0) -- (8.93,4.29) -- cycle    ;
%Straight Lines [id:da0009504312126924486] 
\draw    (439,325.5) -- (390,348.5) ;
\draw [shift={(409.97,339.12)}, rotate = 334.86] [fill={rgb, 255:red, 0; green, 0; blue, 0 }  ][line width=0.08]  [draw opacity=0] (8.93,-4.29) -- (0,0) -- (8.93,4.29) -- cycle    ;
%Straight Lines [id:da8454214741203768] 
\draw    (439,295.5) .. controls (440.67,297.17) and (440.67,298.83) .. (439,300.5) .. controls (437.33,302.17) and (437.33,303.83) .. (439,305.5) .. controls (440.67,307.17) and (440.67,308.83) .. (439,310.5) .. controls (437.33,312.17) and (437.33,313.83) .. (439,315.5) .. controls (440.67,317.17) and (440.67,318.83) .. (439,320.5) .. controls (437.33,322.17) and (437.33,323.83) .. (439,325.5) -- (439,325.5) ;
%Straight Lines [id:da11656615395060521] 
\draw    (439,258.5) -- (439,295.5) ;
\draw [shift={(439,282)}, rotate = 270] [fill={rgb, 255:red, 0; green, 0; blue, 0 }  ][line width=0.08]  [draw opacity=0] (8.93,-4.29) -- (0,0) -- (8.93,4.29) -- cycle    ;
%Straight Lines [id:da5305881809582578] 
\draw    (497,348.5) -- (439,325.5) ;
\draw [shift={(463.35,335.16)}, rotate = 21.63] [fill={rgb, 255:red, 0; green, 0; blue, 0 }  ][line width=0.08]  [draw opacity=0] (8.93,-4.29) -- (0,0) -- (8.93,4.29) -- cycle    ;
%Straight Lines [id:da5661479285567468] 
\draw    (439,295.5) -- (461,296.5) ;
\draw [shift={(454.99,296.23)}, rotate = 182.6] [fill={rgb, 255:red, 0; green, 0; blue, 0 }  ][line width=0.08]  [draw opacity=0] (8.93,-4.29) -- (0,0) -- (8.93,4.29) -- cycle    ;
%Straight Lines [id:da12076032618937815] 
\draw    (439,258.5) .. controls (440.23,256.49) and (441.85,256.09) .. (443.86,257.31) .. controls (445.87,258.53) and (447.49,258.13) .. (448.71,256.12) .. controls (449.94,254.11) and (451.56,253.71) .. (453.57,254.93) .. controls (455.58,256.15) and (457.2,255.75) .. (458.42,253.74) .. controls (459.64,251.72) and (461.26,251.32) .. (463.28,252.54) .. controls (465.29,253.76) and (466.91,253.36) .. (468.14,251.35) .. controls (469.36,249.34) and (470.98,248.94) .. (472.99,250.16) .. controls (475,251.38) and (476.62,250.98) .. (477.85,248.97) .. controls (479.07,246.96) and (480.69,246.56) .. (482.7,247.78) .. controls (484.71,249) and (486.33,248.6) .. (487.56,246.59) -- (492,245.5) -- (492,245.5) ;
%Straight Lines [id:da5043611596786598] 
\draw    (461,296.5) .. controls (463.21,295.68) and (464.72,296.37) .. (465.55,298.58) .. controls (466.37,300.79) and (467.88,301.48) .. (470.09,300.66) .. controls (472.3,299.84) and (473.81,300.53) .. (474.64,302.74) .. controls (475.47,304.95) and (476.98,305.64) .. (479.19,304.82) .. controls (481.4,303.99) and (482.91,304.68) .. (483.74,306.89) .. controls (484.56,309.1) and (486.07,309.79) .. (488.28,308.97) .. controls (490.49,308.15) and (492,308.84) .. (492.83,311.05) -- (496,312.5) -- (496,312.5) ;
%Straight Lines [id:da6529547240928909] 
\draw    (461,296.5) -- (493,284.5) ;
\draw [shift={(481.68,288.74)}, rotate = 159.44] [fill={rgb, 255:red, 0; green, 0; blue, 0 }  ][line width=0.08]  [draw opacity=0] (8.93,-4.29) -- (0,0) -- (8.93,4.29) -- cycle    ;
%Straight Lines [id:da5574411621520772] 
\draw    (551,246.5) -- (601,258.5) ;
\draw [shift={(580.86,253.67)}, rotate = 193.5] [fill={rgb, 255:red, 0; green, 0; blue, 0 }  ][line width=0.08]  [draw opacity=0] (8.93,-4.29) -- (0,0) -- (8.93,4.29) -- cycle    ;
%Straight Lines [id:da8596543561713852] 
\draw    (601,325.5) -- (552,348.5) ;
\draw [shift={(571.97,339.12)}, rotate = 334.86] [fill={rgb, 255:red, 0; green, 0; blue, 0 }  ][line width=0.08]  [draw opacity=0] (8.93,-4.29) -- (0,0) -- (8.93,4.29) -- cycle    ;
%Straight Lines [id:da2271888742964] 
\draw    (601,295.5) .. controls (602.73,293.89) and (604.39,293.95) .. (606,295.68) .. controls (607.6,297.41) and (609.26,297.47) .. (610.99,295.86) .. controls (612.72,294.25) and (614.38,294.31) .. (615.99,296.04) .. controls (617.6,297.76) and (619.27,297.82) .. (620.99,296.21) .. controls (622.72,294.6) and (624.38,294.66) .. (625.98,296.39) -- (629,296.5) -- (629,296.5) ;
%Straight Lines [id:da6869023190300076] 
\draw    (601,258.5) -- (601,295.5) ;
\draw [shift={(601,282)}, rotate = 270] [fill={rgb, 255:red, 0; green, 0; blue, 0 }  ][line width=0.08]  [draw opacity=0] (8.93,-4.29) -- (0,0) -- (8.93,4.29) -- cycle    ;
%Straight Lines [id:da7458813390758704] 
\draw    (657,313.5) -- (629,296.5) ;
\draw [shift={(638.73,302.41)}, rotate = 31.26] [fill={rgb, 255:red, 0; green, 0; blue, 0 }  ][line width=0.08]  [draw opacity=0] (8.93,-4.29) -- (0,0) -- (8.93,4.29) -- cycle    ;
%Straight Lines [id:da16826191275684743] 
\draw    (601,295.5) -- (601,325.5) ;
\draw [shift={(601,315.5)}, rotate = 270] [fill={rgb, 255:red, 0; green, 0; blue, 0 }  ][line width=0.08]  [draw opacity=0] (8.93,-4.29) -- (0,0) -- (8.93,4.29) -- cycle    ;
%Straight Lines [id:da9703865561353382] 
\draw    (601,258.5) .. controls (602.23,256.49) and (603.85,256.09) .. (605.86,257.31) .. controls (607.87,258.53) and (609.49,258.13) .. (610.71,256.12) .. controls (611.94,254.11) and (613.56,253.71) .. (615.57,254.93) .. controls (617.58,256.15) and (619.2,255.75) .. (620.42,253.74) .. controls (621.64,251.72) and (623.26,251.32) .. (625.28,252.54) .. controls (627.29,253.76) and (628.91,253.36) .. (630.14,251.35) .. controls (631.36,249.34) and (632.98,248.94) .. (634.99,250.16) .. controls (637,251.38) and (638.62,250.98) .. (639.85,248.97) .. controls (641.07,246.96) and (642.69,246.56) .. (644.7,247.78) .. controls (646.71,249) and (648.33,248.6) .. (649.56,246.59) -- (654,245.5) -- (654,245.5) ;
%Straight Lines [id:da883066796422478] 
\draw    (601,325.5) .. controls (603.19,324.65) and (604.72,325.32) .. (605.58,327.51) .. controls (606.44,329.7) and (607.97,330.37) .. (610.16,329.52) .. controls (612.35,328.66) and (613.88,329.33) .. (614.74,331.52) .. controls (615.6,333.71) and (617.13,334.38) .. (619.32,333.53) .. controls (621.51,332.68) and (623.04,333.35) .. (623.89,335.54) .. controls (624.75,337.73) and (626.28,338.4) .. (628.47,337.55) .. controls (630.66,336.7) and (632.19,337.37) .. (633.05,339.56) .. controls (633.91,341.75) and (635.44,342.42) .. (637.63,341.57) .. controls (639.82,340.71) and (641.35,341.38) .. (642.21,343.57) .. controls (643.07,345.76) and (644.6,346.43) .. (646.79,345.58) .. controls (648.98,344.73) and (650.51,345.4) .. (651.37,347.59) .. controls (652.23,349.78) and (653.76,350.45) .. (655.95,349.6) -- (658,350.5) -- (658,350.5) ;
%Straight Lines [id:da3205490449698173] 
\draw    (629,296.5) -- (661,280.5) ;
\draw [shift={(649.47,286.26)}, rotate = 153.43] [fill={rgb, 255:red, 0; green, 0; blue, 0 }  ][line width=0.08]  [draw opacity=0] (8.93,-4.29) -- (0,0) -- (8.93,4.29) -- cycle    ;

% Text Node
\draw (24,121.07) node [anchor=north west][inner sep=0.75pt]    {$e^{+}$};
% Text Node
\draw (26,53.4) node [anchor=north west][inner sep=0.75pt]    {$e^{-}$};
% Text Node
\draw (85,79.4) node [anchor=north west][inner sep=0.75pt]    {$\gamma $};
% Text Node
\draw (165,122.4) node [anchor=north west][inner sep=0.75pt]    {$e^{+}$};
% Text Node
\draw (167,93.4) node [anchor=north west][inner sep=0.75pt]    {$e^{-}$};
% Text Node
\draw (164,46.4) node [anchor=north west][inner sep=0.75pt]    {$\gamma $};
% Text Node
\draw (167,71.4) node [anchor=north west][inner sep=0.75pt]    {$\gamma $};
% Text Node
\draw (192,51.4) node [anchor=north west][inner sep=0.75pt]    {$e^{-}$};
% Text Node
\draw (193,118.4) node [anchor=north west][inner sep=0.75pt]    {$e^{+}$};
% Text Node
\draw (333,130.4) node [anchor=north west][inner sep=0.75pt]    {$e^{+}$};
% Text Node
\draw (336,102.4) node [anchor=north west][inner sep=0.75pt]    {$\gamma $};
% Text Node
\draw (334,48.4) node [anchor=north west][inner sep=0.75pt]    {$\gamma $};
% Text Node
\draw (333,76.4) node [anchor=north west][inner sep=0.75pt]    {$e^{-}$};
% Text Node
\draw (357,46.4) node [anchor=north west][inner sep=0.75pt]    {$e^{-}$};
% Text Node
\draw (359,125.4) node [anchor=north west][inner sep=0.75pt]    {$e^{+}$};
% Text Node
\draw (258,69.4) node [anchor=north west][inner sep=0.75pt]    {$\gamma $};
% Text Node
\draw (499,50.4) node [anchor=north west][inner sep=0.75pt]    {$\gamma $};
% Text Node
\draw (498,78.4) node [anchor=north west][inner sep=0.75pt]    {$\gamma $};
% Text Node
\draw (499,131.4) node [anchor=north west][inner sep=0.75pt]    {$e^{+}$};
% Text Node
\draw (498,106.4) node [anchor=north west][inner sep=0.75pt]    {$e^{-}$};
% Text Node
\draw (402,83.4) node [anchor=north west][inner sep=0.75pt]    {$\gamma $};
% Text Node
\draw (528,44.4) node [anchor=north west][inner sep=0.75pt]    {$e^{-}$};
% Text Node
\draw (665,103.4) node [anchor=north west][inner sep=0.75pt]    {$e^{-}$};
% Text Node
\draw (664,132.4) node [anchor=north west][inner sep=0.75pt]    {$\gamma $};
% Text Node
\draw (666,73.4) node [anchor=north west][inner sep=0.75pt]    {$\gamma $};
% Text Node
\draw (665,43.4) node [anchor=north west][inner sep=0.75pt]    {$e^{+}$};
% Text Node
\draw (528,128.4) node [anchor=north west][inner sep=0.75pt]    {$e^{+}$};
% Text Node
\draw (595,43.4) node [anchor=north west][inner sep=0.75pt]    {$\gamma $};
% Text Node
\draw (28,231.4) node [anchor=north west][inner sep=0.75pt]    {$e^{-}$};
% Text Node
\draw (28,331.4) node [anchor=north west][inner sep=0.75pt]    {$e^{+}$};
% Text Node
\draw (160,332.4) node [anchor=north west][inner sep=0.75pt]    {$e^{+}$};
% Text Node
\draw (163,263.4) node [anchor=north west][inner sep=0.75pt]    {$\gamma $};
% Text Node
\draw (163,229.4) node [anchor=north west][inner sep=0.75pt]    {$\gamma $};
% Text Node
\draw (159,295.4) node [anchor=north west][inner sep=0.75pt]    {$e^{-}$};
% Text Node
\draw (197,232.4) node [anchor=north west][inner sep=0.75pt]    {$e^{-}$};
% Text Node
\draw (197,332.4) node [anchor=north west][inner sep=0.75pt]    {$e^{+}$};
% Text Node
\draw (328,259.4) node [anchor=north west][inner sep=0.75pt]    {$e^{+}$};
% Text Node
\draw (331,331.4) node [anchor=north west][inner sep=0.75pt]    {$\gamma $};
% Text Node
\draw (332,228.4) node [anchor=north west][inner sep=0.75pt]    {$\gamma $};
% Text Node
\draw (330,290.4) node [anchor=north west][inner sep=0.75pt]    {$e^{-}$};
% Text Node
\draw (366,237.4) node [anchor=north west][inner sep=0.75pt]    {$e^{-}$};
% Text Node
\draw (366,335.4) node [anchor=north west][inner sep=0.75pt]    {$e^{+}$};
% Text Node
\draw (501,340.4) node [anchor=north west][inner sep=0.75pt]    {$e^{+}$};
% Text Node
\draw (501,304.4) node [anchor=north west][inner sep=0.75pt]    {$\gamma $};
% Text Node
\draw (502,233.4) node [anchor=north west][inner sep=0.75pt]    {$\gamma $};
% Text Node
\draw (499,273.4) node [anchor=north west][inner sep=0.75pt]    {$e^{-}$};
% Text Node
\draw (246,274.4) node [anchor=north west][inner sep=0.75pt]    {$\gamma $};
% Text Node
\draw (81,300.4) node [anchor=north west][inner sep=0.75pt]    {$\gamma $};
% Text Node
\draw (418,299.4) node [anchor=north west][inner sep=0.75pt]    {$\gamma $};
% Text Node
\draw (529,237.4) node [anchor=north west][inner sep=0.75pt]    {$e^{-}$};
% Text Node
\draw (531,338.4) node [anchor=north west][inner sep=0.75pt]    {$e^{+}$};
% Text Node
\draw (662,305.4) node [anchor=north west][inner sep=0.75pt]    {$e^{+}$};
% Text Node
\draw (664,339.4) node [anchor=north west][inner sep=0.75pt]    {$\gamma $};
% Text Node
\draw (664,233.4) node [anchor=north west][inner sep=0.75pt]    {$\gamma $};
% Text Node
\draw (663,271.4) node [anchor=north west][inner sep=0.75pt]    {$e^{-}$};
% Text Node
\draw (609,272.4) node [anchor=north west][inner sep=0.75pt]    {$\gamma $};

\end{tikzpicture}

    \caption{Representative Feynman diagrams for the SM background processes contributing to $e^+ e^- \to e^+ \, e^- \,\gamma \,\gamma$. Additional diagrams, including those with photon exchange in the final state due to identical diphotons, are topologically equivalent to the illustrated cases and are omitted for brevity. }
    \label{fig:SMbackground}
\end{figure}

At a future lepton collider with a center-of-mass energy of $\sqrt{s}=240$~GeV, it is essential to assess the decay length of light ALPs in order to determine whether they decay promptly inside the detector or give rise to displaced or invisible signatures, which directly dictates the appropriate search strategy. The energy distributions and decay lengths for light ALPs ($m_a = 5$ or $10$~GeV) produced in the process $e^+e^- \to e^+e^- a$ are displayed in Fig.~\,\ref{fig:decaylengthenergy}.
The decay length $L_a$, defined as the average distance traveled by the ALP before decaying into photons, depends on its energy $E_a$, mass $m_a$, and total decay width $\Gamma_a$: 
\bea
\Gamma_a & = & \frac{\left( C^2_{\gamma} + \tilde{C}^2_{\gamma} \right) m_a^3 }{4 \pi},\\~~~ 
L_a & = & \frac{E_a}{m_a} \frac{1}{\Gamma_a} = \frac{4\pi E_a}{\left(C_\gamma^2+\tilde{C}_\gamma^2\right)m_a^4},
\eea
where $\tilde{C}_\gamma$ and $C_\gamma$ denote the CP-even and CP-odd couplings, respectively. For weakly coupled ALPs with $\tilde{C}_\gamma,\,C_\gamma=0.01 \ll 1$, the decay length is typically $\mathcal{O}(1)~\mu m$, significantly shorter than the geometric size of detectors~\cite{CEPCStudyGroup:2018ghi}. Consequently, such ALPs are best probed via diphoton final states rather than through missing energy signatures. 
Thus, we focus on the signal process
\begin{equation}
    e^+\,e^-\to e^+\,e^-a\to e^+\,e^-\,\gamma\,\gamma.
\end{equation}
The dominant SM background arises from direct diphoton emission in the process
\begin{equation}
    e^+\,e^-\to e^+\, e^-\,\gamma\,\gamma,
\end{equation}
with representative Feynman diagrams depicted in Fig.~\ref{fig:SMbackground}. Other backgrounds involving misidentified leptons or photons are expected to be subleading after the applied kinematic selections and are neglected in this study.

To optimize the search sensitivity, we categorize the ALP mass range into two distinct regions: $1)$ low-mass region with $5\leq m_a \leq 120$ GeV, and $2)$ high-mass region with $120 < m_a\leq 220$ GeV. 
For ALPs lighter than $5$ GeV, rare decays of heavy hadrons offer superior sensitivity due to enhanced production rates \cite{Agrawal:2021dbo,Jiang:2024cqj}. 
Conversely, for $m_a>220$ GeV, the ALP production cross section at lepton colliders suffers significantly phase-space suppression, reducing detection prospects.

In low-mass region ($5\leq m_a\leq 120$ GeV), both signal and background events are picked up by the following pre-selection cuts in simulation:
\begin{equation}
\begin{aligned}
    p_T^e>10~{\rm GeV},\,|\eta_e|&<2.5,\,\Delta R_{ee}>0.4,\\
    p_T^\gamma>10~{\rm GeV},\,|\eta_\gamma|<2.5&,\,\Delta R_{e\gamma}>0.4,\,\Delta R_{\gamma\gamma}>0.1,
\end{aligned}
\end{equation}
where $p_T^i$ denotes the transverse momentum of particle $i$, $\eta_i$ is the pseudo-rapidity, and $\Delta R_{ij}=\sqrt{( \phi_i-\phi_j )^2+( \eta_i-\eta_j)^2}$ represents the angular separation between particles $i$ and $j$. 
To account for the collinear photon pairs from highly boosted light ALPs, we adopt a relaxed $\Delta R_{\gamma\gamma}(>0.1)$ requirement enabled by one-to-one correspondence reconstruction techniques \cite{Wang:2024eji}.
Further background suppression is achieved through optimized selection cuts summarized in Table.~\,\ref{tab:optimized_small_mass}. 
\begin{table}
\caption{Optimized selection cuts for low-mass ALPs ($5\leq m_a \leq 120$ GeV).}
    \label{tab:optimized_small_mass}
    \centering
    \begin{tabular}{c c c}
    \hline\hline
    Cuts &Description &  Values \\
    \hline 
    1   &Final state selection &$N_e=2,\,N_\gamma=2$ \\
    \hline
    \multirow{2}{*}{2}   &\multirow{2}{*}{Photon kinematics}  &$p_T^{\gamma_1}>20~{\rm GeV},\,p_T^{\gamma_2}>15~{\rm GeV},$\\
        &   &$\,|\eta_\gamma|<1.0,\,\Delta R_{e\gamma}>1.5$ \\ 
    \hline
    3   &VBF-dominant region   &$m_{ee}>100~{\rm GeV},\Delta\eta_{ee}>2.0$ \\
    \hline
    4   &ALP mass windows & $|m_{\gamma\gamma}-m_a|<1~{\rm GeV}$\\
    \hline
    \hline
    \end{tabular}
\end{table}
Here $N_i$ denotes the number of particle $i$, $\Delta\eta_{ee}\equiv\eta_{e^+}-\eta_{e^-}$ is the electron pseudo-rapidity difference, while $m_{ee}$ and $m_{\gamma\gamma}$ represent the electron pair and diphoton invariant masses, respectively. 
The superscripts $\gamma_1$ and $\gamma_2$ denote leading and sub-leading photon ordered by transverse momentum $p_T$. 
Table.~\,\ref{tab:Cutflow_small_mass} presents the cutflow for signal (assuming $\tilde{C}_\gamma=C_\gamma=0.1$) and background processes at benchmark mass points. 
The stringent requirements on the photon with small pseudo-rapidity $|\eta_\gamma|$, large transverse momentum $p_T^\gamma$ and a large electron-photon separation $\Delta R_{e\gamma}$ effectively suppress SM backgrounds where the radiative photons are typically soft and collinear to electrons. 
Additionally, the VBF-dominant selection ($m_{ee}>100$ GeV, $\Delta\eta_{ee}>2.0$) and the narrow diphoton mass window ($|m_{\gamma\gamma}-m_a|<1$ GeV) provide a significant signal-background discrimination. 
\begin{table}
\caption{Cutflow for low-mass ALP signals (assuming $\tilde{C}_\gamma=C_\gamma=0.1$) and SM backgrounds.}
    \label{tab:Cutflow_small_mass}
\resizebox{1.0\columnwidth}{!}{
    \centering
    \begin{tabular}{c|c|c|c}
    \hline
    \hline
    \multirow{2}{*}{Cuts} &\multicolumn{3}{c}{Cross section for signal (background) [pb] } \\
    \cline{2-4}
    &$m_a=5$ GeV &$m_a=50$ GeV &$m_a=90$ GeV\\
    \hline
    Pre-selection &$3.09\times10^{-3}~~(0.082
    )$ &$3.77\times10^{-3}~~(0.082)$  &$2.54\times10^{-3}~~(0.082)$ \\
    \hline
    Cut $1$ &$2.73\times10^{-3}~~(0.058)$  & $3.31\times10^{-3}~~(0.058)$  &$2.20\times10^{-3}~~(0.058)$ \\
    \hline
    Cut $2$ &$1.25\times10^{-3}~~(1.67\times10^{-3})$   &$1.14\times10^{-3}~~(1.67\times10^{-3})$  &$7.42\times10^{-4}~~(1.67\times10^{-3})$ \\
    \hline
    Cut $3$ &$8.76\times10^{-4}~~(2.61\times10^{-4})$   &$8.43\times10^{-4}~~(2.61\times10^{-4})$   &$5.05\times10^{-4}~~(2.61\times10^{-4})$ \\
    \hline
    Cut $4$ &$8.73\times10^{-4}~~(2.10\times10^{-6})$   &$7.20\times10^{-4}~~(8.31\times10^{-6})$   &$3.47\times10^{-4}~~(2.41\times10^{-6})$ \\
    \hline
    \hline
    \end{tabular}
    }
\end{table}

For the high-mass region ($120<m_a\leq220$ GeV), we apply the following pre-selection cuts to both signal and background events: 
\begin{equation}
\begin{aligned}
    p_T^e>10~{\rm GeV},\,|\eta_e|&<2.5,\,\Delta R_{ee}>0.4,\\
    p_T^\gamma>10~{\rm GeV},\,|\eta_\gamma|<2.5&,\,\Delta R_{e\gamma}>0.4,\,\Delta R_{\gamma\gamma}>0.4.
\end{aligned}
\end{equation}
The optimized selection strategy for this mass region is summarized in Table.~\,\ref{tab:optimized_large_mass}, with the correspondence cutflow presented in Table.~\,\ref{tab:Cutflow_large_mass}. 
In this mass region, the higher ALP energy allows stricter photon transverse momentum requirements to effectively suppress SM background. 
To mitigate phase-space suppression of the signals, we remove the $m_{ee}$ cut while maintaining $\Delta\eta_{ee}>1.0$ to select signals arising from VBF process. 
Accounting for the degraded energy resolution at higher photon energies, we expand the ALP mass reconstruction windows to $\pm2$ GeV around $m_a$. 
These optimized selections maintains high signal efficiency while achieving significant background rejection, as demonstrated by the cutflow results showing three or four orders of magnitude background suppression across all mass points. 

\begin{table}
\caption{Optimized selection cuts for for high-mass ALPs ($120 < m_a \leq 220$ GeV).}
    \label{tab:optimized_large_mass}
    \centering
    \begin{tabular}{c c c}
    \hline\hline
    Cuts &Description &  Values \\
    \hline 
    1   &Final state selection &$N_e=2,\,N_\gamma=2$ \\
    \hline
    \multirow{2}{*}{2}   &\multirow{2}{*}{Photon kinematics}  &$p_T^{\gamma_1}>40~{\rm GeV},\,p_T^{\gamma_2}>25~{\rm GeV},$\\
        &   &$\,|\eta_\gamma|<1.0,\,\Delta R_{e\gamma}>1.0$ \\ 
    \hline
    3   &VBF-dominant region   &$\Delta\eta_{ee}>1.0$ \\
    \hline
    4   &ALP-mass window  & $|m_{\gamma\gamma}-m_a|<2~{\rm GeV}$\\
    \hline
    \hline
    \end{tabular}
\end{table}
\begin{table}
\caption{Cutflow for high-mass ALP signals (assuming $\tilde{C}_\gamma=C_\gamma=0.1$) and SM backgrounds. 
In the scenario where $m_a = 125$ GeV, the background arising from $e^+e^-\to e^+e^-h\,(\to\gamma\gamma)$ channel has been involved in our results. }
    \label{tab:Cutflow_large_mass}
\resizebox{1.0\columnwidth}{!}{
    \centering
    \begin{tabular}{c|c|c|c}
    \hline
    \hline
    \multirow{2}{*}{Cuts} &\multicolumn{3}{c}{Cross section for signal (background) [pb] } \\
    \cline{2-4}
    &$m_a=125$ GeV &$m_a=170$ GeV &$m_a=200$ GeV\\
    \hline
    Pre-selection   &$1.42\times10^{-3}~~(0.081)$   &$3.85\times10^{-4}~~(0.081)$   &$6.19\times10^{-5}~~(0.081)$   \\
    \hline
    Cut $1$    &$1.16\times10^{-3}~~(0.057)$   &$3.06\times10^{-4}~~(0.057)$   &$4.60\times10^{-5}~~(0.057)$ \\
    \hline
    Cut $2$   &$5.81\times10^{-4}~~(1.40\times10^{-3})$   &$1.46\times10^{-4}~~(1.41\times10^{-3})$   &$1.88\times10^{-5}~~(1.41\times10^{-3})$   \\
    \hline
    Cut $3$    &$4.96\times10^{-4}~~(4.50\times10^{-4}
    )$   &$1.10\times10^{-4}~~(4.51\times10^{-4})$    &$9.91\times10^{-6}~~(4.51\times10^{-4})$    \\
    \hline
    Cut $4$    &$4.76\times10^{-4}~~(1.32\times10^{-5})$   &$9.61\times10^{-5}~~(1.71\times10^{-6})$   &$7.97\times10^{-6}~~(3.24\times10^{-7})$   \\
    \hline
    \hline
    \end{tabular}
    }
\end{table}

\subsection{Likelihood Analysis}

\begin{figure}
\centering
\includegraphics[width=0.49\linewidth]{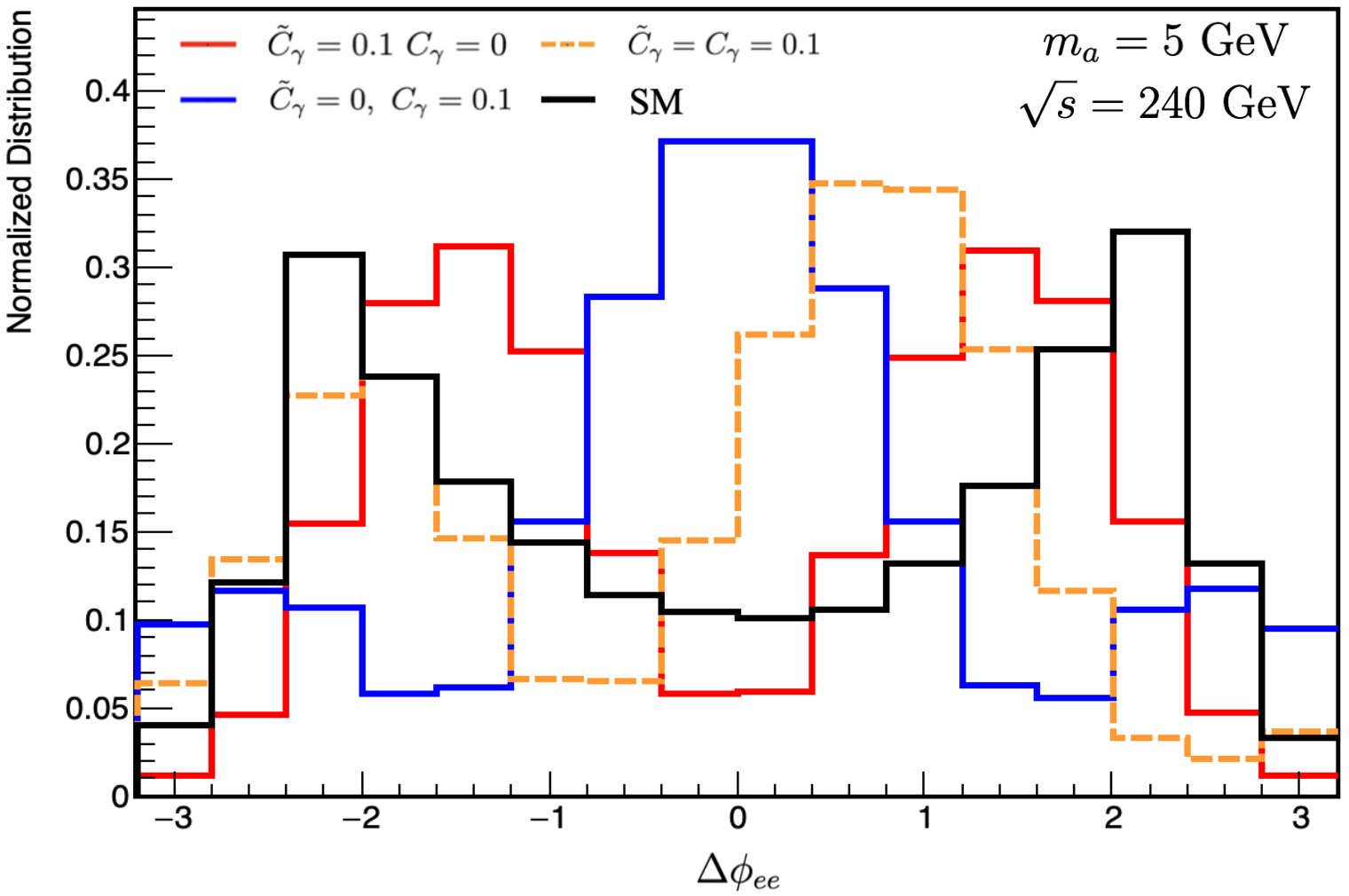}
\includegraphics[width=0.487\linewidth]{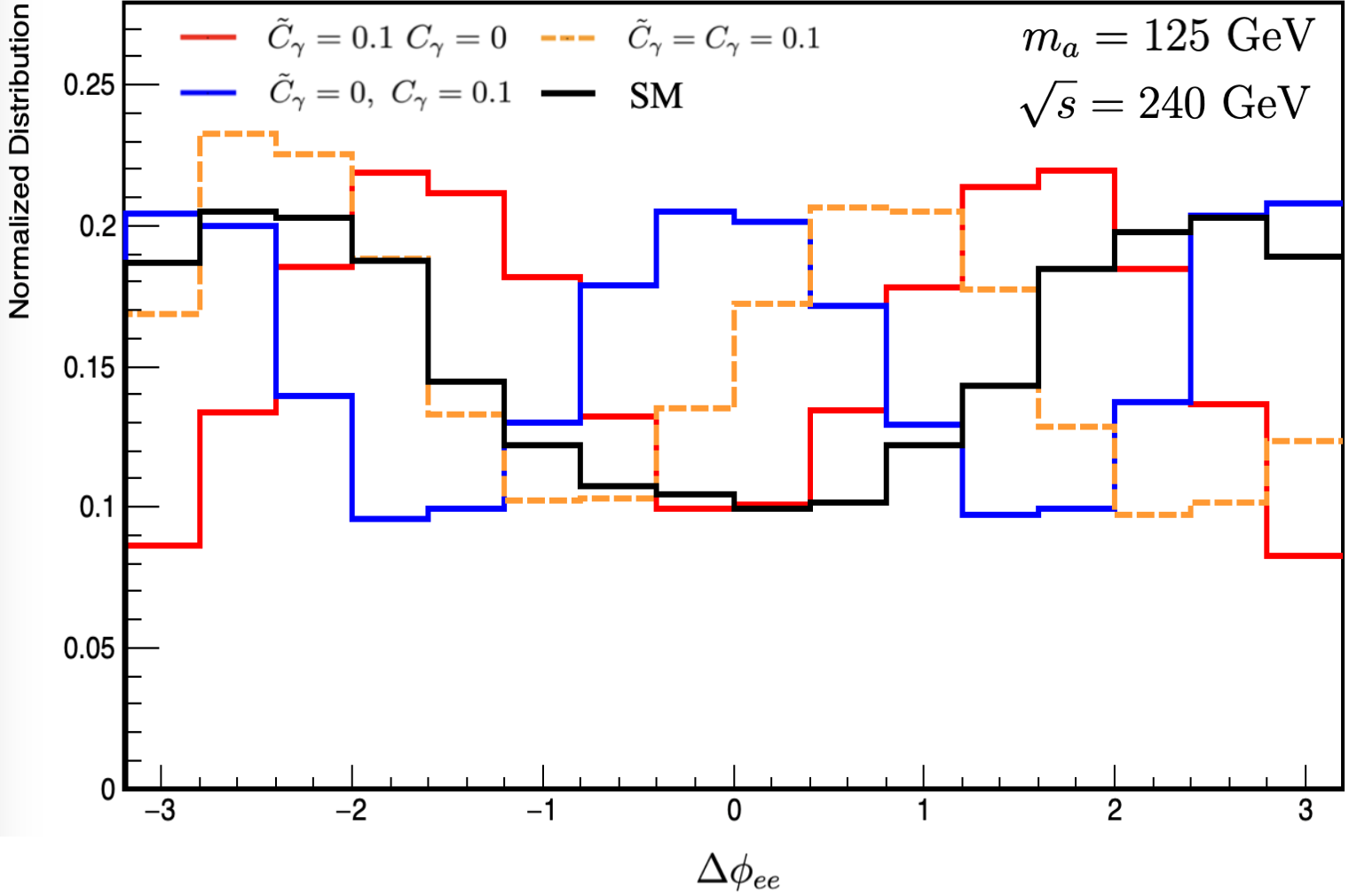}
\caption{Normalized differential distributions of $\Delta\phi_{ee}$ for ALP masses $m_a=5$ GeV (left panel) and $125$ GeV (right panel). 
The black curves represent the SM background.
The red and blue curves correspond to the contribution from the CP-conserving ($aF_{\mu\nu}\tilde{F}^{\mu\nu}$) and CP-violating ($aF_{\mu\nu}F^{\mu\nu}$) ALP-photon interaction.
The orange curves denote the coexistence of two equal couplings, i.e.,$\tilde{C}_\gamma=C_\gamma$. 
}
\label{fig:DeltaPhiee_eeAA_Fiducial}
\end{figure}

Within the kinematic selection criteria outlined in the previous subsection, the normalized differential distributions of the observable $\Delta\phi_{ee}$ for both signal and background processes are presented in Fig.~\,\ref{fig:DeltaPhiee_eeAA_Fiducial}.
Because the CP information is encoded in the shape—rather than the overall normalization—of the $\Delta\phi_{ee}$ distribution, this observable provides a powerful discriminator for identifying ALP-induced effects.
To fully exploit this feature, we perform a binned likelihood analysis based on the $\Delta\phi_{ee}$ distribution, following the formalism in Ref.\,\cite{Cowan:2010js}, in order to optimize the sensitivity to ALP signals. 
For a meaningful comparison with existing limits from $e$EDM experiments \cite{ACME:2018yjb,Caldwell:2022xwj,Roussy:2022cmp}, we defined the $90\%$ C.L. exclusion significance $\sigma_{\text{exc}}$ as:
\begin{equation}
\begin{aligned}
\sigma_{\text{exc}} &= \sqrt{-2\,\ln \left( \frac{L(S+B| B)}{ L (B|B)} \right) } \geq 1.65, 
\end{aligned} \label{equ:likelihood_exclusion}
\end{equation}
where $L$ denotes the Poisson likelihood, $B$ is the expected number of background events, and $S$ is the number of signal events. 
In our analysis, the signal yield is decomposed as $S=\tilde{S}_{\gamma}+S_{\gamma}+S_{\tilde{\gamma}\gamma}$, where $\tilde{S}_\gamma$ arises from the CP-conserving ALP-photon interaction and scales with $\tilde{C}_\gamma^2$, $S_\gamma$ originates from the CP-violating ALP-photon interaction and scales with $C_\gamma^2$, and $S_{\tilde{\gamma}\gamma}$ corresponds to their interference, scaling with $\tilde{C}_\gamma C_\gamma$. 
The binned Poisson likelihood function is given by:
\bea
L(X | Y) = \prod_i^N \frac{x_i^{y_i}}{y_i!} \, e^{-x_i}, 
\eea
where $x_i$ is the predicted event number under hypothesis $X$ in the $i$-th bin of the $\Delta\phi_{ee}$ distribution, and $y_i$ is the corresponding assumed observed number of events under hypothesis $Y$.

To complement the exclusion analysis discussed above and to explore the discovery potential of ALPs at future lepton colliders, we further define the discovery significance as 
\begin{equation}
\begin{aligned}
\sigma_{\text{dis}} &= \sqrt{-2\,\ln \left( \frac{L(B | S+B)}{ L (S+B|S+B)} \right) } \geq 5,
\end{aligned} \label{equ:likelihood_discovery}
\end{equation}
where a value of $\sigma_{\text{dis}}\geq5$ corresponds to the conventional $5\sigma$ threshold for discovery. 
Beyond discovery and exclusion, the CP-nature dynamics of the ALP can subsequently be probed through kinematic observables. 
Leveraging the strong discriminatory power of the $\Delta\phi_{ee}$ observable, particularly its sensitivity to the CP structure of the ALP-photon interaction, we extend the likelihood analysis to determine the dominant CP component of the signal.
Assuming that an ALP signal has been established, we test whether the observed $\Delta\phi_{ee}$ distribution can be consistently described by a single CP structure, namely a purely CP-conserving or purely CP-violating interaction, using
\begin{equation}
\begin{aligned}\label{equ:likelihood_CPV}
\sigma_{\tilde{C}_\gamma}&=\sqrt{-2\,\ln\left( \frac{L(\tilde{S}_{\gamma}+B|S+B)}{L(S+B|S+B)} \right)}\leq 2,~&\text{with $95\%$ C.L.},
\end{aligned}
\end{equation}
\begin{equation}
\begin{aligned}\label{equ:likelihood_CPV2}
\sigma_{C_\gamma}&=\sqrt{-2\,\ln\left( \frac{L(S_\gamma+B|S+B)}{L(S+B|S+B)} \right)}\leq 2, ~&\text{with $95\%$ C.L}.
\end{aligned}
\end{equation}
where $\sigma_{\tilde{C}_\gamma}$ corresponds to the region where CP-even ALP-photon interaction dominates the differential distribution in $\Delta\phi_{ee}$, and $\sigma_{C_\gamma}$ represents the region for CP-odd interaction.

\section{Results and Discussion}\label{sec:result}

\subsection{Constraints from future lepton colliders and low-energy measurements. }

\begin{figure}
\centering
\hspace{-0.4cm}
\includegraphics[width=0.49\linewidth]{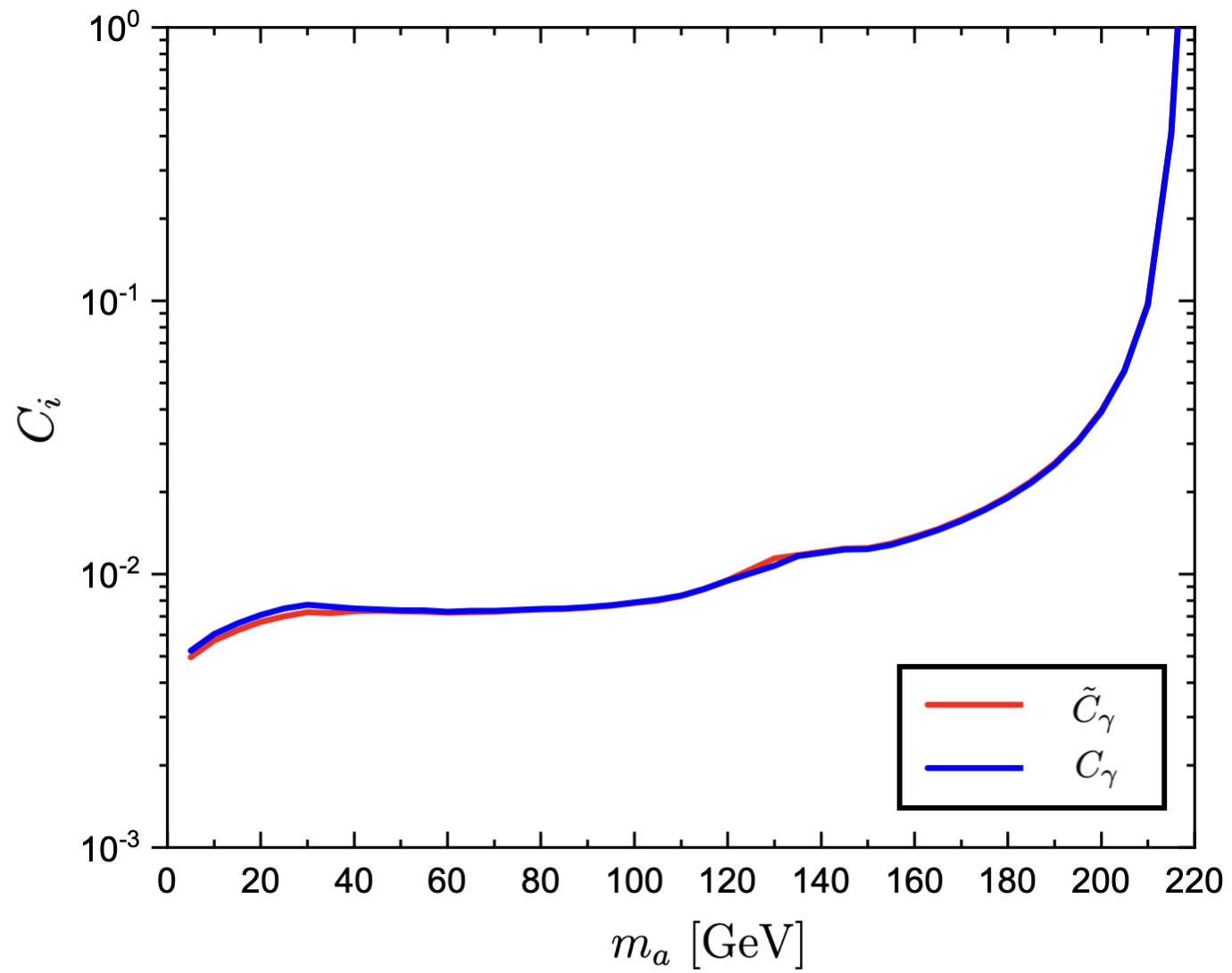}
\hspace{0.1cm}
\includegraphics[width=0.49\linewidth]{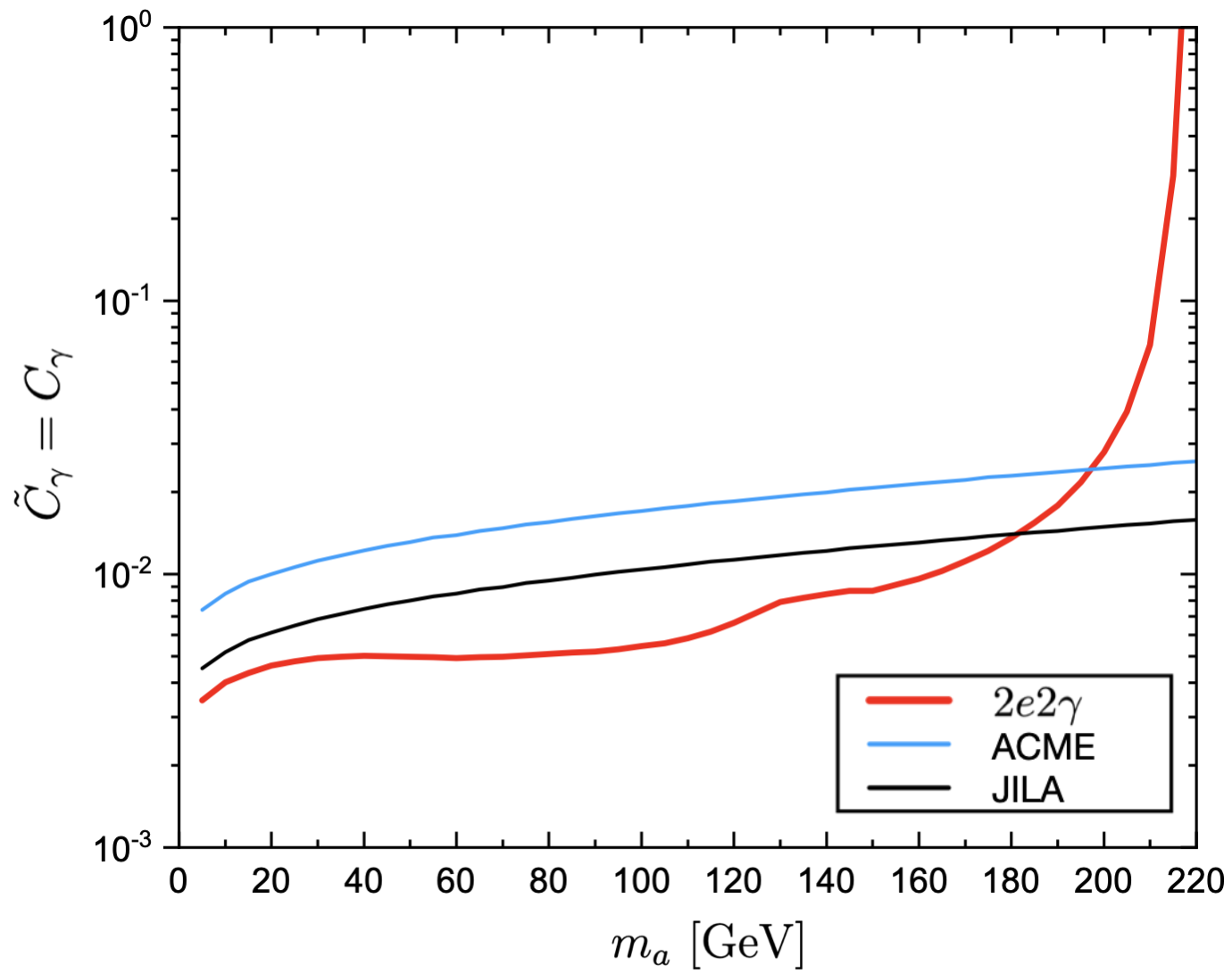}
\caption{Expected $90\%$~C.L. constraints on the ALP-photon coupling constants at a lepton collider with collison energy of 240 GeV and integrated luminosity of 5 ab$^{-1}$. 
\textbf{Left panel:} constraints assuming only one type of ALP-photon interaction is present-either the CP-conserving operator $aF_{\mu\nu}\tilde{F}^{\mu\nu}$  or the CP-violating operators $aF_{\mu\nu}F^{\mu\nu}$. 
\textbf{Right panel:} constraints under the assumption of equal coupling strengths, i.e., $\tilde{C}_\gamma=C_\gamma$. 
The bounds from the $e$EDM measurements by the ACME collaboration \cite{ACME:2018yjb} and the JILA collaboration \cite{Caldwell:2022xwj,Roussy:2022cmp} are also shown for comparison.
}
\label{fig:Constraint}
\end{figure}

Based on the phenomenological analysis presented in the previous section, Fig.\,\ref{fig:Constraint} shows the expected constraints on the ALP-photon coupling constants at a future lepton collider operating at $\sqrt{s}=240~$GeV with an integral luminosity $\mathcal{L}=5$~ab$^{-1}$. 
In scenarios with a single operator--either the CP-conserving or CP-violating ALP-photon interaction--future lepton colliders can constrain the corresponding coupling constants ($\tilde{C}_\gamma$ or $C_\gamma$) down to $\mathcal{O}(10^{-2})$. 
By contrast, when both operators contribute with equal strength ($\tilde{C}_\gamma=C_\gamma$), the sensitivity improves to $\mathcal{O}(10^{-3})$ over a wide range of ALP masses. 
This projected sensitivity already surpasses the current indirect limits derived from $e$EDM measurements~\cite{ACME:2018yjb,Caldwell:2022xwj,Roussy:2022cmp}, highlighting the strong potential of direct collider searches.

\begin{figure}[t!]
\centering
\includegraphics[width=0.45\linewidth]{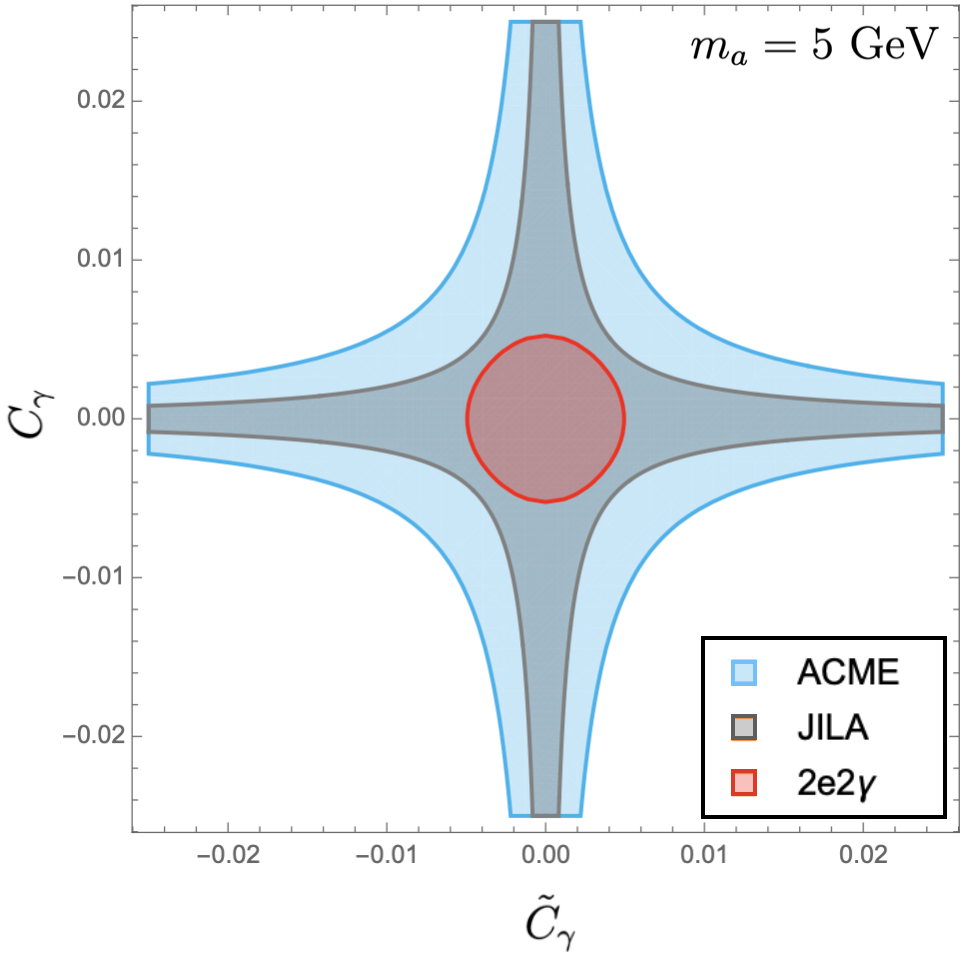}
\hspace{0.2cm}
\includegraphics[width=0.45\linewidth]{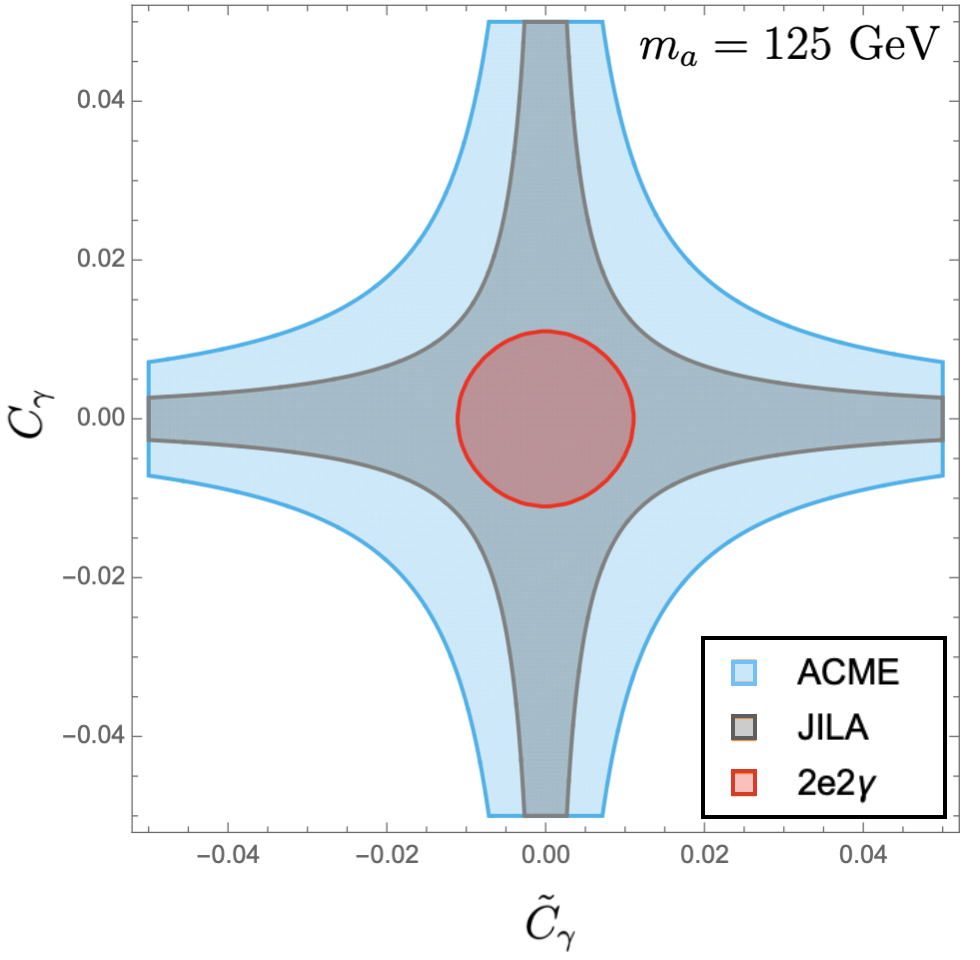}
\caption{Two-dimensional constraints on the ALP-photon couplings for $m_a=5~$GeV (left) and $m_a=125~$GeV (right) at a lepton collider with collison energy of 240 GeV and integrated luminosity of 5 ab$^{-1}$. 
The blue and black shaded regions show the parameter space allowed by the ACME and JILA $e$EDM measurements, respectively. 
The red shaded areas indicate the sensitivity accessible at future lepton colliders. 
}
\label{fig:2dim_constraints}
\end{figure}

To further explore the general scenario in which both couplings coexist without assuming any specific relation between them, we perform a two-dimensional analysis in the $(\widetilde{C}_\gamma, C_\gamma)$ parameter space. The resulting sensitivity contours are shown in Fig.~\ref{fig:2dim_constraints}.
These results demonstrate that the projected sensitivities at future lepton colliders are substantially stronger than those inferred from $e$EDM data. 
This underscores the advantage of direct collider searches in setting robust and comprehensive limits on CP violation ALP-photon interaction.

\subsection{Discriminating the CP structure of ALP interactions}

\begin{figure}
\centering
\hspace{-0.2cm}
\includegraphics[width=0.49\linewidth]{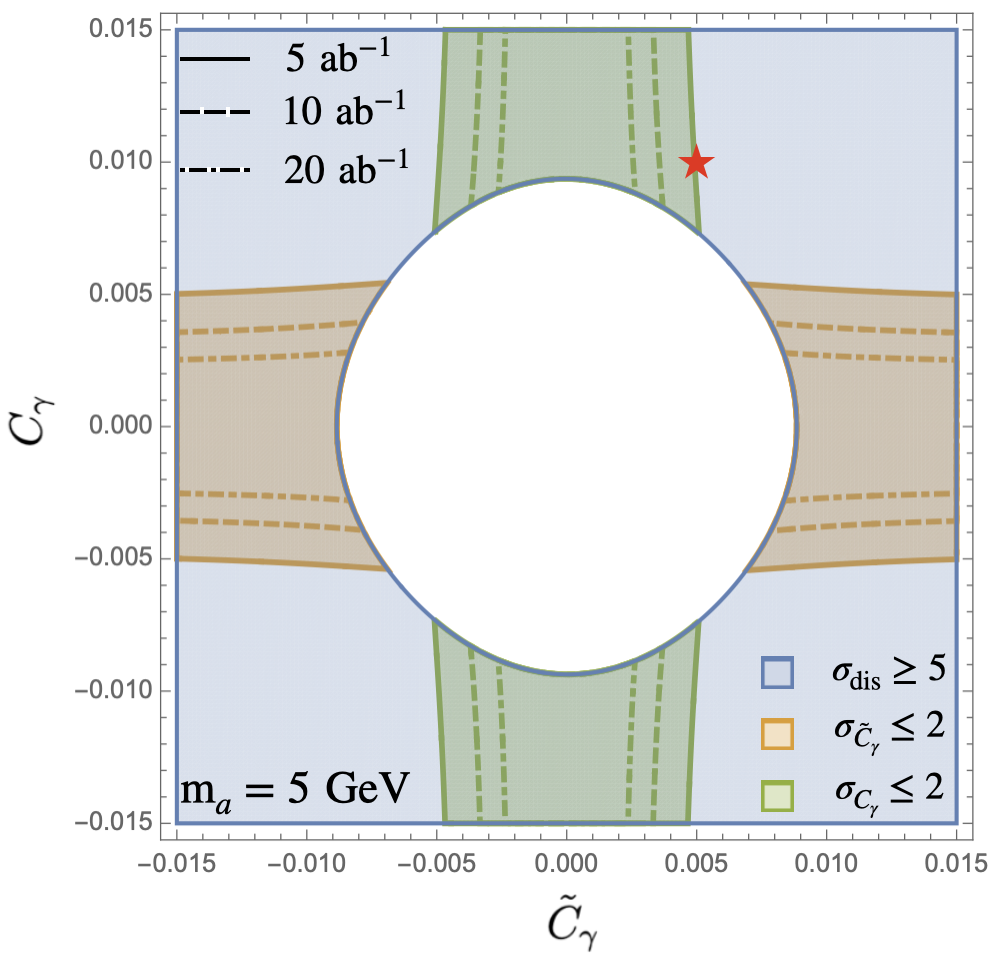}\hspace{0.1cm}
\includegraphics[width=0.487\linewidth]{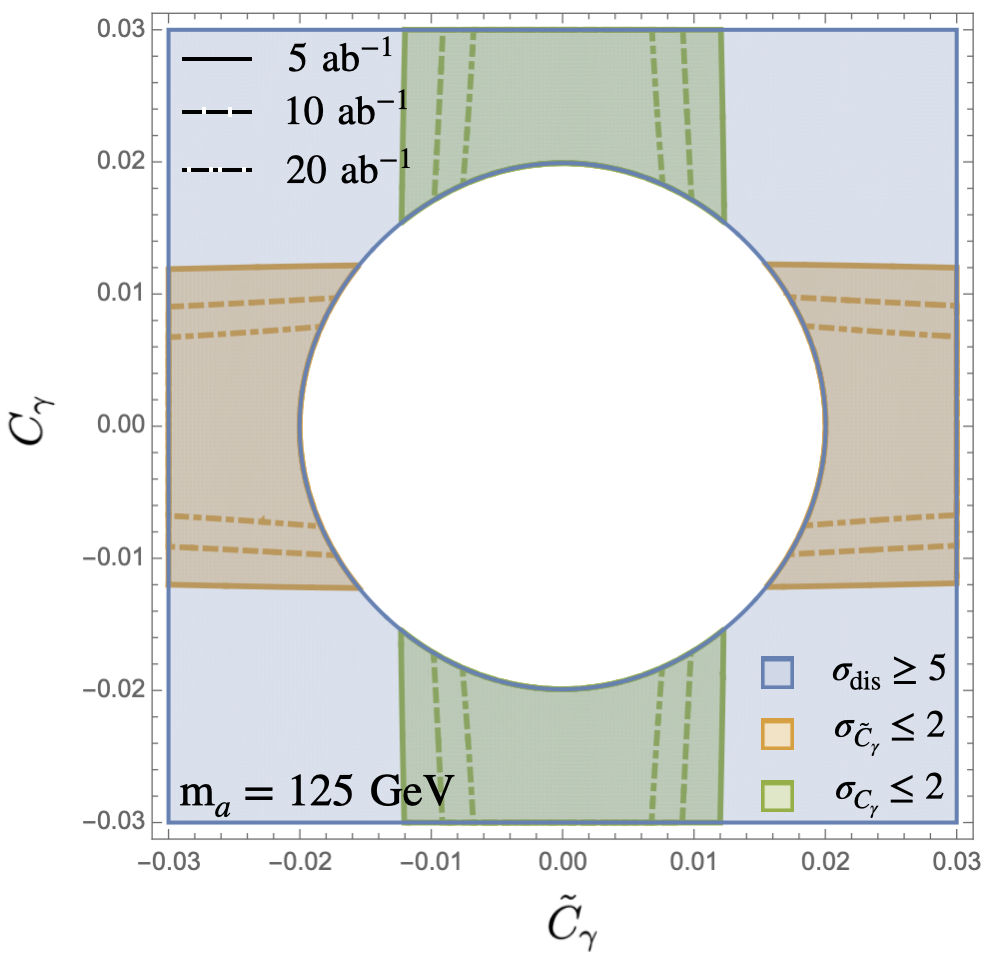}
\caption{
Parameter regions for ALP discovery and CP-structure identification at a lepton collider with $\sqrt{s}=240$ GeV. 
The blue shaded areas indicate the $5\sigma$ discovery reach for the ALP signal with an integrated luminosity of $5~\mathrm{ab}^{-1}$. 
The orange (green) shaded regions denote parameter points where the observed $\Delta\phi_{ee}$ distribution remains compatible at $95\%$ C.L. with a single-operator interpretation, corresponding to the CP-even operator $aF_{\mu\nu}\widetilde{F}^{\mu\nu}$ (CP-odd operator $aF_{\mu\nu}F^{\mu\nu}$), respectively.
Regions inside the discovery reach but outside both the orange and green areas correspond to parameter space where neither single-operator hypothesis can describe the observed $\Delta\phi_{ee}$ distribution. In these regions, the data require the simultaneous presence of both operators, providing direct collider evidence for CP violation in the ALP–photon interaction.
The solid, dashed, and dot-dashed curves show the corresponding compatibility boundaries for integrated luminosities of $5$, $10$, and $20~\mathrm{ab}^{-1}$, respectively. 
The red star marks a benchmark point $(\widetilde{C}_\gamma=0.005,\; C_\gamma=0.01)$ illustrating how increased luminosity improves the ability to resolve interference effects between the two operators.
}
\label{fig:luminosity}
\end{figure}

A major advantage of a future lepton collider is that, once an ALP signal is observed, it can further determine the CP structure of the ALP--photon interaction using the shape information of the $\Delta\phi_{ee}$ distribution. In particular, $\Delta\phi_{ee}$ cleanly separates the pure CP-even and CP-odd hypotheses, and it is also sensitive to the interference contribution when both operators are simultaneously present, as shown in Fig.\,\ref{fig:DeltaPhiee_eeAA_Fiducial}. 
The results of this CP-structure discrimination are summarized in Fig.\,\ref{fig:luminosity} for cases $m_a=5$ GeV and 125 GeV. The blue shaded regions denote the $5\sigma$ discovery reach, defined by the test statistic in Eq.\,\eqref{equ:likelihood_discovery}. The orange (green) shaded regions correspond to parameter points where the observed $\Delta\phi_{ee}$ distribution remains compatible at $95\%$ C.L. with a single-operator interpretation, namely a purely CP-even (CP-odd) hypothesis, quantified by Eqs.\,\eqref{equ:likelihood_CPV} and \eqref{equ:likelihood_CPV2}, respectively.
This indicates that, within the orange (green) shaded regions, the observed $\Delta\phi_{ee}$ distribution cannot exclude the purely CP-even (CP-odd) hypothesis at the $95\%$ C.L., and therefore the presence of the CP-odd (CP-even) operator cannot be statistically established.

The clear separation between the orange and green regions indicates that signals dominated by a single CP structure can be reliably distinguished, avoiding confusion between the CP-even and CP-odd interaction hypotheses. However, more interestingly, there exist parameter regions where a $5\sigma$ signal is discoverable (blue) but neither single-operator hypothesis is compatible with the observed $\Delta\phi_{ee}$ distribution at 95\% C.L. These regions are concentrated around $|\widetilde{C}_\gamma|\sim|C_\gamma|$, where the interference contribution induces a characteristic distortion of the $\Delta\phi_{ee}$ shape that cannot be reproduced by either operator alone. In this case, the data require the simultaneous presence of both operators, implying a CP-violating ALP--photon interaction within this effective description.

It is also noteworthy that increasing the integrated luminosity significantly improves the sensitivity to such interference effects. As an illustrative example, the benchmark point marked by the red star corresponds to $\widetilde{C}_\gamma = 0.005$ and $C_\gamma = 0.01$. With an integrated luminosity of $5~\mathrm{ab}^{-1}$, this point remains compatible with the CP-odd-only hypothesis at the $95\%$ C.L. However, increasing the luminosity to $10~\mathrm{ab}^{-1}$ substantially shrinks the parameter region allowed by the CP-odd-only hypothesis and excludes this benchmark point. This demonstrates that the observed $\Delta\phi_{ee}$ distribution cannot be explained by the operator $aF_{\mu\nu}F^{\mu\nu}$ alone and instead requires the simultaneous presence of $aF_{\mu\nu}\widetilde{F}^{\mu\nu}$.

\section{Conclusion}\label{sec:conclu}

The search for axion-like particles (ALPs) and the exploration of their CP properties represent a critical frontier in probing physics beyond the Standard Model. In this study, we investigated the potential of future lepton colliders operating at $\sqrt{s}=240$ GeV with an integrated luminosity of $\mathcal{L}=5.0~\mathrm{ab}^{-1}$ to probe both the production signal and the CP structure of ALP–photon interactions through the dimension-5 operators $aF_{\mu\nu}\tilde{F}^{\mu\nu}$ (CP-even) and $aF_{\mu\nu}F^{\mu\nu}$ (CP-odd). Although such couplings are conventionally constrained by low-energy observables such as the electron electric dipole moment ($e$EDM), our analysis demonstrates the unique capability of high-energy colliders not only to probe these interactions directly, but also to determine their CP nature through the shape of kinematic observables such as the azimuthal angle difference $\Delta\phi_{ee}$.

In this work we focused on the production channel $e^+e^- \to e^+e^- a$, where the ALP is predominantly produced via vector boson fusion (VBF). Given the short decay length of the ALP relative to detector coverage, we reconstruct the full $e^+e^-\gamma\gamma$ final state. The clean environment at future lepton colliders provides an ideal setting to suppress Standard Model backgrounds, primarily from QED $\gamma\gamma$ processes, and isolate the ALP signal. To probe the ALP–photon couplings, we analyzed the azimuthal angle difference $\Delta\phi_{ee}$ between the final-state electrons, which is sensitive to the CP structure of the ALP interactions. Four event-selection strategies were developed for two mass regions: a low-mass region with $5 \le m_a \le 120$ GeV and a high-mass region with $120 < m_a \le 220$ GeV. A binned log-likelihood fit to the reconstructed $\Delta\phi_{ee}$ distributions was then performed to statistically constrain the couplings $\tilde{C}_\gamma$ and $C_\gamma$. Compared with low-energy precision experiments such as the $e$EDM, this collider-based approach provides complementary access to the ALP parameter space through direct production. We find that future lepton colliders can achieve stronger constraints than those currently obtained from $e$EDM measurements.

Beyond setting constraints, our analysis demonstrates the potential to resolve the CP nature of the ALP–photon interaction. The shape of the $\Delta\phi_{ee}$ distribution allows one to distinguish between CP-even and CP-odd dominant scenarios once a signal is observed, while simultaneously revealing the characteristic interference pattern that arises when both operators are present. As shown in Fig.~8, the regions dominated by CP-even and CP-odd interactions are clearly separated, whereas the presence of both operators leads to distinctive signatures that cannot be reproduced by either interaction alone. In particular, for benchmark scenarios where $|\tilde{C}_\gamma| \simeq |C_\gamma|$, the observed distribution falls outside the regions compatible with a single-operator hypothesis, providing direct evidence for CP-violating effects. Furthermore, increasing the integrated luminosity significantly improves the sensitivity to such interference patterns, especially when one coupling is subdominant. This establishes a powerful collider-based method to uncover the CP structure of ALPs, offering a direct probe of new sources of CP violation beyond the Standard Model.

In summary, our analysis demonstrates that future lepton colliders provide a powerful and complementary probe of ALP–photon interactions beyond the reach of current low-energy measurements. While eEDM experiments place stringent indirect constraints on CP-violating couplings, collider observables enable a direct and model-independent test of the ALP interaction structure. In particular, the azimuthal angle difference $\Delta\phi_{ee}$ between the final-state electrons serves as a robust CP-sensitive observable that distinguishes CP-even, CP-odd, and mixed interaction scenarios through the shape of its differential distribution. Our results show that future lepton colliders operating at $\sqrt{s}=240$ GeV with integrated luminosities of a few ab$^{-1}$ can probe ALP–photon couplings down to $\mathcal{O}(10^{-3})$, exceeding the current limits from $e$EDM measurements. More importantly, when both CP-conserving and CP-violating operators are present with comparable strengths, the interference pattern in the $\Delta\phi_{ee}$ distribution enables a direct identification of CP violation in the ALP sector. This demonstrates that future lepton colliders can not only discover ALPs but also determine the CP nature of their interactions, providing a unique window into the structure of new physics beyond the Standard Model.

\acknowledgments

We thank Manqi Ruan and Yong-Feng Zhu for their helpful discussion. The work of Y.L. is partly supported by the National Science Foundation of China under Grant Nos. 12075257, 12175016 and the National Key R$\&$D Program of China under Grant No. 2023YFA1607104. The work of MYS is partly supported by Grants South China Normal University Young Teachers Scientific Research Foundation (No.599/672203).

\bibliographystyle{jhep}{}
\bibliography{sample.bib}

\end{document}